%% file: main.tex
\documentclass[apj,numberedappendix]{emulateapj}

\usepackage[sort&compress]{natbib}
\bibpunct{(}{)}{;}{a}{}{,}
\usepackage{color}
\definecolor{darkgreen}{RGB}{0,142,128}

\usepackage{hyperref}
\hypersetup{colorlinks,citecolor=blue,linkcolor=blue}

\usepackage{amsmath}
\usepackage{amsthm}
\usepackage{amsfonts}
\usepackage{amssymb}
\usepackage{url}
\usepackage{subfigure}
\usepackage{multirow}

\newcommand{\refff}[1]{{#1}}
\newcommand{\reff}[1]{{#1}}
\newcommand{\modifAS}[1]{{#1}}
\newcommand{\modifASt}[1]{{#1}}
\newcommand{\modifPC}[1]{{#1}}

\newcommand{\lrsp}{LRSP}
\newcommand{\jfm}{\textit{JFM}}
\newcommand{\gafd}{\textit{Geophys. Astrophys. Fluid Dyn.} }
\newcommand{\adv}{\textit{Adv. Space Res.}}

\begin{document}

\title{On the sensitivity of magnetic cycles in global simulations of solar-like stars}
\shorttitle{On the sensitivity of magnetic cycles in global simulations of solar-like stars}

\author{A. Strugarek}
\affil{Laboratoire AIM Paris-Saclay, CEA/Irfu Universit\'e Paris-Diderot CNRS/INSU, F-91191 Gif-sur-Yvette.}
\affil{D\'epartement de physique, Universit\'e de Montr\'eal, C.P. 6128 Succ. Centre-Ville, Montr\'eal, QC H3C-3J7, Canada}
\email{antoine.strugarek@cea.fr}
\author{P. Beaudoin}
\affil{D\'epartement de physique, Universit\'e de Montr\'eal, C.P. 6128 Succ. Centre-Ville, Montr\'eal, QC H3C-3J7, Canada}
\author{P. Charbonneau}
\affil{D\'epartement de physique, Universit\'e de Montr\'eal, C.P. 6128 Succ. Centre-Ville, Montr\'eal, QC H3C-3J7, Canada}
\author{A. S. Brun}
\affil{Laboratoire AIM Paris-Saclay, CEA/Irfu Universit\'e Paris-Diderot CNRS/INSU, F-91191 Gif-sur-Yvette.}

\shortauthors{A. Strugarek, et al.}
\begin{abstract}
The periods of magnetic activity cycles in the Sun and solar-type
stars do not exhibit a simple or even single trend with respect to
rotation rate or luminosity. Dynamo models can be used to interpret
this diversity, and can ultimately help us understand why some
solar-like stars do not exhibit a magnetic cycle, whereas some do, and
for the latter what physical mechanisms set their magnetic cycle
period. Three-dimensional non-linear \modifPC{magnetohydrodynamical
  simulations} present the advantage of having only a small number of
tunable parameters, and produce in a dynamically self-consistent
manner the flows and the dynamo magnetic fields pervading stellar
interiors. We conducted a series of such simulations within the
EULAG-MHD framework, varying the rotation rate and luminosity of the
modeled solar-like convective envelopes. We find decadal magnetic
cycles when the Rossby number near the base of the convection zone is
moderate (typically between 0.25 and 1). Secondary, shorter cycles
located at the top of the convective envelope close to the equator are
also observed in our numerical experiments, when the local Rossby
number is lower than 1. The \reff{deep-seated} dynamo sustained in these numerical experiments is fundamentally non-linear, in that it is the feedback of the large-scale magnetic field on the large-scale differential rotation that sets the magnetic cycle period. The cycle period is found to decrease with the Rossby number, which offers an alternative theoretical explanation to the variety of activity cycles observed in solar-like stars.
\end{abstract}


\maketitle

\section{Introduction}
\label{sec:introduction}

\def\Prot{P_{\rm rot}}
\def\Pcyc{P_{\rm cyc}}
\def\Ro{\rm Ro}
\def\tauc{\tau_{\rm c}}
Over fifty years ago, Olin Wilson initiated the Mt Wilson
stellar activity monitoring program relying on a measure
of emission in the core of the
Calcium H$+$K lines. This remarkable observational
effort was further extended by other
similar observational programs, some still ongoing
(see \citealt{Hall:2008cu} and references therein).
Spatially-resolved solar observations
had already long shown that emission in the Ca H$+$K lines
is closely associated
with surface magnetism, in particular active regions and surrounding
plages. A large fraction of the solar-type stars in the Mt Wilson
sample were found to exhibit temporal variability in H$+$K emission, often in
the form of more or less regular defined cycles, with periods ranging from
5 to some 20 yr
(\citealt{Wilson:1978is}; \citealt{Baliunas:1995cc}; see also \citealt{Hall:2007iq}). Such cyclic activity is mostly
observed in stars of masses, rotation rates
and evolutionary states similar to the Sun. The natural
interpretation is to consider these as stellar analogs
of the 11-yr solar magnetic activity cycle. However, even amongst solar-type
stars in the sample, some show highly
irregular, non-cyclic variability or very little variability over
decades; magnetism
is clearly ubiquitous among solar-type stars, but
regular activity cycles are not.

Efforts at understanding these stellar activity cycles have been
hampered by the fact that most stars in which cycles have been measured
are field stars, for which the precise evolutionary status (age,
luminosity, etc.) was difficult to pin down accurately. This situation
has changed in the past decade, following accurate parallax
determination by space missions such as Hipparcos and GAIA,
and in some cases by asteroseismic analyses made possible
by photometric observations from the Corot and Kepler missions \citep[\textit{e.g.}][]{Mathur:2012bj,Chaplin:2014jf,doNascimento:2014is}.

Again quite naturally, physical
interpretation of observed stellar cycles has been sought
within solar dynamo theory
(e.g. \citealt{Noyes:1984bp,1996ApJ...460..848B,Saar:1999dg}).
Working on 13 Mt-Wilson stars with well-defined cycle periods, \citet{Noyes:1984bp}
used simple mean-field dynamo models to establish empirical relationship linking 
rotation period ($\Prot$), cycle period ($\Pcyc$), and spectral
type. The Rossby number ($\Ro=\Prot/\tauc$), a dimensionless quantity defined
as the ratio of rotation period to the convective turnover time ($\tauc$),
emerged from their analysis as the key parameter. From the dynamo
point of view this is satisfying, as in this context the Rossby number
measures the influence of rotation on convection. Large-scale differential rotation 
and cyclonic turbulence, in turn, are two key inductive processes in
solar dynamo theory, and both require this influence (see, e.g., 
\citealt{Miesch:2009kb,Charbonneau:2014dy} and references therein).
\citet{Noyes:1984bp} showed that cycle periods in their reduced stellar sample could be
tolerably well-represented by the relationship $\Pcyc\propto\Ro^{5/4}$;
yet their modeling work also showed that any such relationship
is sensitive to modeling choices made in setting up the dynamo
model.

In a similar vein, but working with an expanded sample and
longer datasets, \citet{Saar:1999dg} showed that the significant
scatter about the \citet{Noyes:1984bp} empirical relationship could be
markedly reduced by first dividing the sample into ``active'' and ``inactive''
stars (based on the overall emission levels), leading to two
``branches'' in the $\Pcyc$ vs $\Ro$ diagram. In their analyses
both branches showed an increase of $\Pcyc$ with $\Ro$,
but with different slopes. \citet{BohmVitense:2007bd}
further suggested an evolutionary scenario whereby
each branch is associated with a distinct
dynamo mode within the star, a proposal buttressed by the fact
that some stars showed dual-period cyclic variability, with the two
periods each lying on one of the two activity branches. As they age
and spin down, young
rapidly rotating stars initially populating the active
branch eventually switch to a less efficient dynamo mode,
and transit to the inactive branch. This scenario, plausible
and compatible with observed stellar activity cycles, remains
unsatisfactory in one important aspect: the Sun, arguably the
best data point of the lot, is left hanging far in between the two stellar
activity branches, forcing one to assume that our star just happens
to be in its transition phase between the two branches
(for a new-and-improved version of this type of scenario, see
\citealt{Metcalfe:2017kx}). Larger samples of stars with a well-defined cyclic activity are today emerging, 
suggesting that this gap between the branches is actually a selection bias and questioning the rationale behind this interpretation \citep[\textit{e.g.}, see][]{BoroSaikia:2018wj}. 

The use of dynamo models to interpret stellar activity cycles,
as pioneered by \citet{Noyes:1984bp}, requires a number of input
``ingredients''. Broadly defined and at minimum: (1) Choosing
a specific class of solar/stellar dynamo model; (2) Choosing
a nonlinear amplitude-limiting mechanism; (3) Specifying how
relevant inductive flows (differential rotation, meridional
circulation, cyclonic turbulence) vary as a function of
rotation rate, structural evolution, etc. Even in the solar
case there is currently no consensus on the first two items
(see, e.g., discussions in \citealt{Charbonneau:2010bf,Brun:2017dy}).
For the present
day Sun the differential rotation (point 3) is well-constrained by helioseismology.
However,
in the case of stars, where comparable asteroseismic
constraints are difficult to obtain, even under physically reasonable
assumptions there remains sufficient leeway to produce
almost any $\Pcyc$ vs $\Prot$ relationship
(see, e.g., \citealt{Charbonneau:2001uh,Jouve:2010jl}).

Global magnetohydrodynamical (MHD) simulations of solar/stellar convection
and associated dynamo action in principle bypass these problems.
Available computational resources limit the range of spatial and temporal
scales that can be captured by such simulations, with the consequence
that some potentially important physical mechanisms
cannot be included, for example the surface decay of active
regions. 
\reff{Nevertheless}, the generation of large-scale flows
by rotationally-influenced convection as well as
the non-linear backreaction of magnetism on
internal flows are both captured in a dynamically correct manner at all
resolved scales. 

As a case in point, the transition from ``solar-like'' differential
rotation (equatorial regions rotating faster than the poles) to
``anti-solar'' differential rotation (slower equator) is now
well-understood. Not surprisingly, the Rossby number emerges again
as a key parameter, with high-$\Ro$ ($\gtrsim 1$) convection leading
to anti-solar differential rotation, and intermediate-$\Ro$ ($\sim 0.1$)
convection producing solar-like differential rotation
(see \citealt{Guerrero:2013hb,Gastine:2014jr,Featherstone:2015bv,Mabuchi:2015ee,Brun:2017em}).
Interestingly, these simulation results also indicate that the Sun may
be near the tipping point between these two rotational
regimes. \reff{It must be noted however that the limited range of spatial scales in
  numerical simulations may lead one to over-estimate the amplitude of
convective flows (see \citealt{Hanasoge:2015ij,Greer:2015ff} and references therein), which would in turn
change the critical stellar Rossby number at which this transition
occur. 
}
One \refff{could nevertheless} infer from these simulation results that a
simple, monotonic relationship between cycle and rotation periods is
not to be expected, considering that differential
rotation is a key inductive process in the vast majority of
extant solar/stellar dynamo scenarios.

Finding counterparts of stellar magnetic cycles in global MHD
simulations has proven a challenging undertaking. While thermally-driven
turbulent convection has no difficulty producing magnetic fields
(e.g. \citealt{Brun:2004ji}), their solar-like organization into a
well-structured and cyclically
varying large-scale component has only been achieved relatively
recently \citep{Ghizaru:2010im,Brown:2011fm,Racine:2011gh,Nelson:2011hw,Nelson:2013fa,Passos:2014kx,Fan:2014ct,Masada:2013fc,Kapyla:2012dg,Kapyla:2016hg,Augustson:2015er}. There are actually few adjustable physical parameters
in such simulations once a background solar/stellar structure is set:
typically luminosity and rotation are the two primary physical knobs,
and both are very well-constrained in the solar case. While low-Ro simulations tend to produce more cyclic solution, the overall variety
of cyclic (and non-cyclic) behaviors uncovered in the afore-cited simulations
likely reflects, at least in part,
the key influence of small-scale dissipative processes,
which are handled quite differently across computational frameworks.

In a recent paper \citep{Strugarek:2017go}, we have presented a set
of global MHD simulations of solar convection produced using
the EULAG-MHD framework (see \citealt{Smolarkiewicz:2013hq,Strugarek:2016bk} and references therein),
with the rotation rates varying between 0.55 and 1.1 times solar, and
the convective luminosity between 0.2 and 0.61 solar luminosity.
Most simulations in the set exhibit fairly regular
large-scale magnetic cycles with decadal-type
periods, the latter showing systematic variations with rotation rate
and luminosity.
\reff{We revisit the dependency of cycle periods
on stellar characteristics with this simulation set}. A striking result from our analyses
is the {\it decrease} of the magnetic cycle period with increasing rotation
period. Indeed, cycles in the simulation set are well represented
by the relationship
$P_{\rm cyc} \propto {\rm Ro}^{-1.8}$; here the negative exponent implies
a trend opposite to that suggested by \citet{Noyes:1984bp}.
A key aspect of this new trend is the
dependency of cycle period on luminosity; when accounted for
in estimating Rossby numbers for the Mt Wilson stellar sample,
all solar twins with well-defined periods are back
on a single branch in the $\Pcyc$--$\Ro$ diagram.
More importantly perhaps, the Sun is now on the same branch,
finally back amongst other solar-type stars.

\modifAS{In this paper we extend the Rossby number range of the \citet{Strugarek:2017go}
simulation set, and show the existence of three well-defined dynamo regimes as a function of the Rossby number: low-Ro with short magnetic cycles appearing close to the surface at low latitude; intermediate-Ro where deep-seated solar-like cycles are observed; and high-Ro where stable wreaths of large-scale magnetic field develop (see Fig. \ref{fig:SummaryPlot} for a summary).} We further carry out a detailed analysis of the nonlinear
driving of large-scale magnetic polarity reversals by magnetically-mediated
alterations of the internal differential rotation.

The remainder of the paper is structured as follows. We present the numerical EULAG-MHD model on which this study is based in \S\ref{sec:conv-zone-model}. 
\modifAS{The dynamo states achieved in the series of 17 simulations analyzed in this paper are summarized in \S\ref{sec:brief-overview}. One prototype cyclic solution is then detailed in \S\ref{sec:prot-cycl-solut}, 
and the non-linear dynamo acting in these simulations is presented in \S\ref{sec:dynamo-action}. The trends of the magnetic cycles periods are finally discussed in \S\ref{sec:trends-scaling-laws}, and we conclude in \S\ref{sec:conclusions}.}

\section{Turbulent convection zone model}
\label{sec:conv-zone-model}

Building on the preliminary
work of \citet{Strugarek:2016bk}, we consider again here spherical fully convective shells with a solar-like
aspect ratio (the inner and outer radii are defined by
$R_{\rm bot}=0.7\,R_\odot$ and $R_{\rm top}=R_\odot$). We detail in what follows the set of anelastic equations
we use to describe the turbulent state of the convective layers
of solar-like stars (\S\ref{sec:form-anel-equat}), and the background
and ambient state around which the aforementioned perturbed equations
are derived (\S\ref{sec:backgr-ambi-stat}). We also explain the
modelling technique used to drive turbulent
convective motions as well as boundary conditions in
\S\ref{sec:numerical-methods}. 

\subsection{Formulation of the anelastic equations}
\label{sec:form-anel-equat}

We consider the Lantz-Braginsky-Roberts
\citep[LBR, see][]{Lantz:1999hm,Braginsky:1995kd} ---or,
equivalently, the Lipps-Hemler \citep{Lipps:1982km,Lipps:1985kg}--- set of anelastic equations
(see \citealp{Vasil:2013ij} for a review on various approximations of anelastic equations).
The perturbed equations are written with respect to an ambient state
(hereafter denoted by the subscript $a$) that theoretically may
differ from the background state (hereafter denoted with overbars) around
which the generic anelastic equations
are derived. The
background state is chosen to be isentropic, and both the ambient and
background states are assumed to satisfy hydrostatic equilibrium.
The anelastic equations written in the stellar rotating frame $\boldsymbol{\Omega}_\star$ are
\begin{align}
 \label{eq:LBR}
 \nabla\cdot\left(\bar{\rho}\mathbf{u}\right) & =  0 \, , \\
  \label{eq:LBRET_u}\mbox{D}_{t} \mathbf{u} =& -\nabla
  \left(\frac{p}{\bar{\rho}}\right) - \frac{\Theta}{\bar{\Theta}}\mathbf{g} -
  2\boldsymbol{\Omega}_\star\times\mathbf{u} +
                                               \frac{\left(\nabla\times{\bf
                                               B}\right)\times{\bf B}}{\mu_0\bar{\rho}} \, , \\
  \label{eq:LBRET_T}\mbox{D}_{t} \Theta =& 
-\left(\mathbf{u}\cdot\nabla\right)\Theta_{a} - \frac{\Theta}{\tau}
\, , \\
  \label{eq:LBR_B}
\mbox{D}_t {\bf B} =&  \left({\bf B}\cdot \nabla\right){\bf u} -
                      \left(\nabla\cdot{\bf u}\right){\bf B}\, ,
\end{align}
where the perturbed quantities are denoted without prime for the sake
of simplicity, and $\mbox{D}_{t}$ is the
material derivative. We recall that we use standard notation for the
basic quantities, \textit{i.e.} ${\bf u}$ is the fluid velocity,
$\rho$ its density, $p$ its pressure, $\Theta$ its potential temperature, and
${\bf B}$ is the magnetic field.  In addition, we use a standard perfect gas
equation of state which is
linearized around the background state. 

\reff{In the preceding equations,
convection is forced by the combined advection of the
unstable ambient entropy profile $S_a$, and a Newtonian cooling term of
characteristic timescale $\tau$ \citep[for
details,
see][]{Prusa:2008df,Smolarkiewicz:2013hq}. The Newtonian cooling damps
entropy perturbations over the timescale
$\tau$ which is always chosen to exceed the convective overturning
time (here, $\tau = 600$ days). It ensures that on long time-scales, the model mimics a stellar
convection zone remaining in thermal equilibrium
\citep[\textit{e.g.}][]{Cossette:2017kf}. The volumetric forcing
  of convection is an alternative to forcing via an imposed heat flux
  across the boundaries of the computational domain. Volumetric forcing models directly the
divergence of the
  diffusive flux \refff{across the domain}, and can be shown to
be mathematically
  and physically
  equivalent to boundary forcing, provided the ambiant state
is adequately specified (see
  Appendix A.1 in \citealt{Cossette:2017kf}). \refff{A
    relaxed state is then achieved, which results from the equilibrium
    between the convective heat transport and the volumetric forcing
    towards the prescribed ambient state.} 
The
  convective luminosity ends up being set by different control
  parameters under each of these two modelling approaches,
hence any direct comparison must be carried out with care.}


\subsection{Background and ambient states}
\label{sec:backgr-ambi-stat}

Our numerical setup closely follows the anelastic benchmark of
\citet{Jones:2011in}. We consider a spherical shell of aspect ratio $\beta=R_{\rm top}/R_{\rm bot}$, and
we set $d = R_{\rm top}-R_{\rm bot} = R_{\rm top}(1-\beta)$. In all simulations presented here, the aspect ratio is solar-like, \textit{i.e.} $\beta=0.7$.
We assume a gravity profile $g=GM/r^{2}$, for which the
anelastic equations admit an equilibrium (denoted by overbars)
polytropic solution \citep[see, \textit{e.g.},][]{Jones:2011in}
\begin{align}
  \label{eq:polytrop}
  \bar{\rho} = \rho_{i}\xi^{n}, \,\, \bar{P} &=
  P_{i}\xi^{n+1},\,\,  \bar{T} = T_{i}\xi\, , \\ \xi &= c_{0} +
                                                       \frac{c_{1}d}{r}\, ,
  \label{eq:polytrop2}
\end{align}
where $n$ is the polytropic index, $\rho_i, P_i, T_i$ are the
density, pressure and temperature at the bottom of the domain, and the constants $c_{0}$ and
$c_{1}$ are given by
\begin{align}
  \label{eq:csts}
  c_{o} =& \frac{2\alpha-\beta-1}{1-\beta} , \, \, c_{1} =
  \frac{(1+\beta)(1-\alpha)}{(1-\beta)^{2}},\\
  \mbox{with } \alpha =& \frac{\beta+1}{\beta \exp\left(N_{\rho}/n\right)+1} \, ,
\end{align}
and where $N_{\rho}=\ln(\rho_{i}/\rho_{o})$ is the number of density scale
heights in the layer. The density at the base of the domain is fixed to $\rho_i=200$ kg/m$^3$ in all the models.
The background potential temperature is given by
\begin{align}
  \label{eq:entropy}
  \bar{\Theta} = \frac{\bar{P}^{1/\gamma}}{\bar{\rho}} = \frac{P_i^{1/\gamma}}{\rho_i}\xi^{(n+1-n\gamma)/\gamma} \, ,
\end{align}
where the standard adiabatic
exponent for a perfect gas is $\gamma=c_{p}/c_{v}=5/3$.
We chose a polytropic exponent $n=3/2$ to naturally ensure an
isentropic background state with $\bar{\Theta} =
P_i^{1/\gamma}/\rho_i$, and the equivalent background entropy profile is $\bar{S} = c_p \ln \left( \bar{\Theta}\right)$ (we consider here $c_p = 3.4 \times 10^{4}$ J/kg/K).

The ambient state needs to be specified only in terms of 
potential temperature. The entropy jump throughout the domain, $\Delta
S$, is used to define the ambient entropy profile by
\begin{align}
  S_{a}(r) = \bar{S} + \Delta S \frac{\xi^{-n}(R_{\rm top}) -
    \xi^{-n}(r)}{\xi^{-n}(R_{\rm top}) - \xi^{-n}(R_{\rm bot})}\, ,
\end{align}
which we recall is related to the ambient potential temperature profile by
$\Theta_a = \exp{\left(S_a/c_p\right)}$.
 

\subsection{Numerical method}
\label{sec:numerical-methods}

The Eulerian-Lagrangian (EULAG) code is designed to use either Eulerian (flux form) or semi-Lagrangian
(advective form) integration schemes \citep[see][]{Prusa:2008df,Smolarkiewicz:2013hq}.
In the case presented here, Eqs. (\ref{eq:LBR}-\ref{eq:LBR_B})
are written as a set of Eulerian conservation laws and projected on a
spherical coordinate system \citep{Prusa:2003fa}. 
EULAG solves the evolution equations
using MPDATA (multidimensional positive definite advection transport 
algorithm), which belongs to the class of nonoscillatory Lax-Wendroff 
schemes \citep{Smolarkiewicz:2006dz}, and is more specifically a second-order-accurate
nonoscillatory forward-in-time template. Since all dissipation is delegated to MPDATA,
this provides an implicit turbulence model
\citep{Domaradzki:2003cb}. 
In EULAG, all linear forcing terms are
integrated in time using a second-order Crank-Nicholson scheme.

In terms of boundary conditions, we consider
stress-free, impermeable boundaries at top and bottom of
the domain such that
\begin{align}
  \label{eq:v_boundaries}
  u_{r} = 0 \, ; \, \partial_{r} \left(u_{\theta}/r\right) = \partial_{r} \left(
    u_{\varphi}/r\right) = 0
\end{align}
at both the top and bottom boundaries. The potential temperature gradient is set to zero on the upper and lower
boundaries. The magnetic boundary conditions are chosen to be perfect conductor at the bottom and purely radial field at the top, such that 
\begin{align}
  \label{eq:b_boundaries_1}
  B_{r} = 0 \, ; \, \partial_{r} \left(r B_{\theta}\right) = \partial_{r} \left(
    r B_{\varphi}\right) = 0\, \mbox{ at } r=R_{\rm bot}\, ,  \\ 
  \label{eq:b_boundaries_2}
    B_\theta = B_\varphi = 0 \mbox{ at } r=R_{\rm top}\, .
\end{align}
The simulations presented in this work are initialized with
random, small-amplitude perturbations around the background profiles
defined in \S\ref{sec:backgr-ambi-stat}.

\modifAS{
The originality of this numerical method lies in the ILES approach
used in the EULAG-MHD code. While this was shown to minimize
dissipation at large-scales, it also renders comparison with other
simulation results less straightforward. Following
\citet{Strugarek:2016bk}, we characterized our simulations with
effective dissipation coefficients deduced from a spectral analysis of
the simulations results. The details of this technical procedure are
given in Appendix \ref{sec:DissipativeProp}. The numerical dissipation
of the various MHD fields fit a classical laplacian operator very well
for the smallest scales of the simulation (spherical harmonics $l >
25$, typically). We find that the effective dissipation coefficients
are of the order of $10^{8}$ m$^2$/s, which is what is typically
expected for the relatively coarse numerical resolution adopted in
this work \citep{Strugarek:2016bk}. The Prandtl number $Pr = \nu_{\rm
  eff}/\kappa_{\rm eff}$ and magnetic Prandtl number $Pm = \nu_{\rm
  eff}/\eta_{\rm eff}$ are generally slightly larger than one,
\reff{which is what can be expected from such a numerical
  approach when using the same subgrid-scale model for all quantities \citep{Domaradzki:2003cb}}. Comparison with other
simulation results must be carried out with care, though, as the
dissipation coefficients reported in Table \ref{ta:dissipation} (see
Appendix \ref{sec:DissipativeProp}) characterize well the small scales
in the simulations, but are by no means representative of the
dissipation of the largest spatial scales associated with the magnetic
cycles.} \refff{Moreover, locally the dissipation can be
 significantly different than that of a Laplacian operator (for more on this see \citealt{Strugarek:2016bk}).}

\modifAS{

\section{Overview of the simulation set}
\label{sec:brief-overview}

\subsection{Dynamo regimes in the simulation set}
\label{sec:dynamo-regimes}

We present here a series of 17 simulations using the same numerical
grid and varying the rotation rate, entropy contrast and number of
density scale heights of the convection zone. The parameters of the
models are listed in Table \ref{ta:Params}. All the simulations presented here exhibit either a solar-like differential rotation self-sustained by convective turbulence under the influence of rotation \citep[e.g.][]{Brun:2002gi}, or an anti-solar differential rotation when the Rossby number reaches values higher than 1 (see panel h in Fig. \ref{fig:3repcases} and \citealt{Brun:2017em}). Several dynamo states are achieved in this series with distinctive properties as shown in Fig. \ref{fig:3repcases}. Some of the simulations display a magnetic cycle deep-seated in the convective envelope, such as model O2 (panel a). A short cycle located close to the equator and in subsurface layers is also observed in some simulations, such as model O6 (panel d2). Finally, in some simulations a stable wreath of magnetic field builds in the bottom part of the convection zone, showing only little time variability due to the convective motions perturbing it \citep[see also][]{Brown:2010cn,Nelson:2013fa} . For instance this is the case of model s01, shown in the bottom panels. Notice also that this particular case possesses an anti-solar differential rotation profile. Finally, some models (not shown here, see \textit{e.g.} Fig. \ref{fig:RossbyTrendSec}) also present stochastic reversals of the deep-seated magnetic field.

\begin{figure*}[htb]
  \centering
  \includegraphics[width=\linewidth]{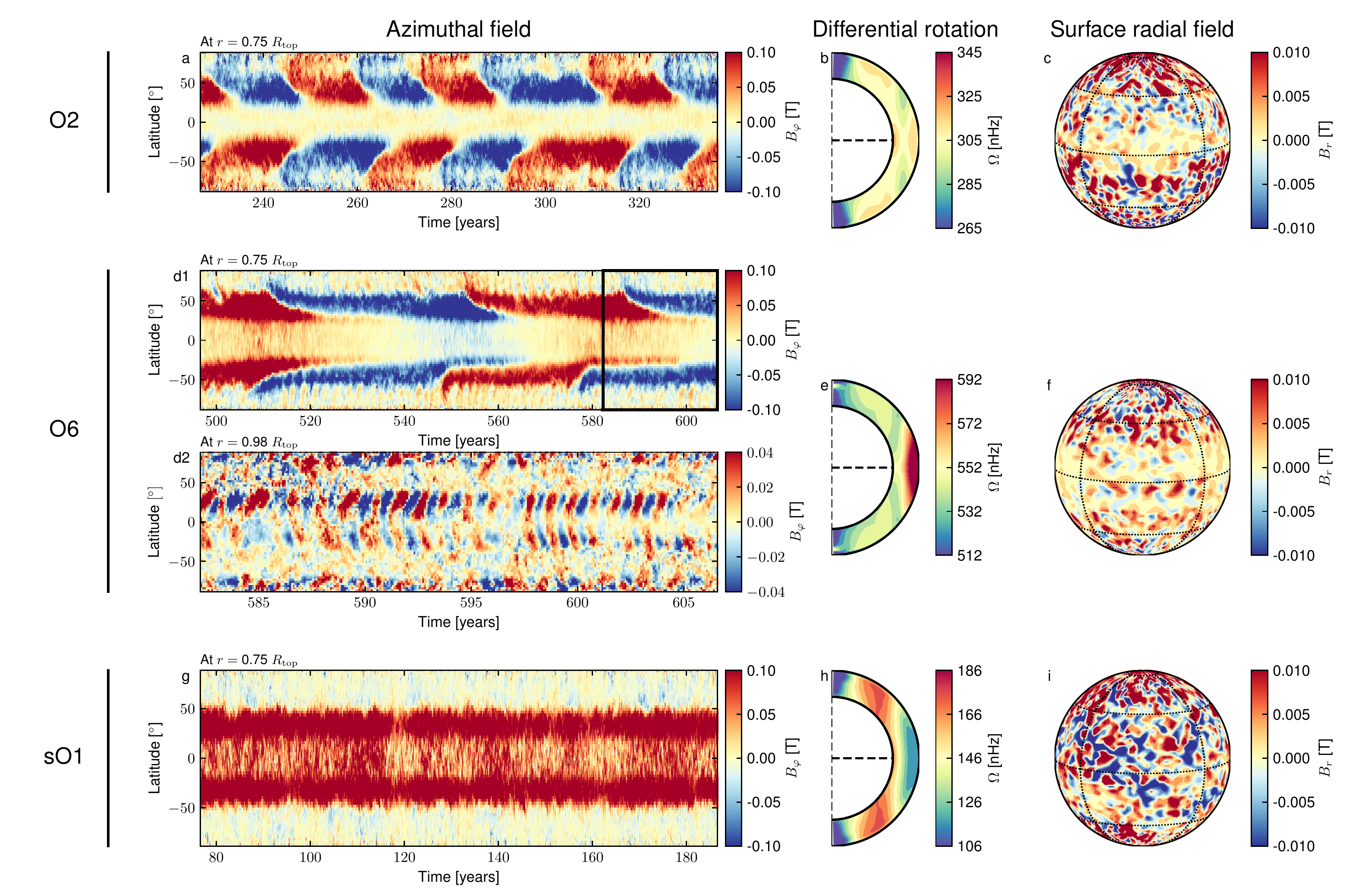}
  \caption{Three representative cases of the dynamo states achieve in this work. From left to right, we represent the azimuthal magnetic field averaged over longitude $\left\langle B_\varphi\right\rangle_{\varphi}$ at $r=0.75 R_{\rm top}$ as a function of latitude and time; the differential rotation profile averaged over hundreds of years; and an instantaneous radial magnetic field in an orthographic projection close to the top boundary of the domain. Panels a to c represent a typical cyclic solution (model O2). Panels d1, e and f represent model 06, and in panel d2 we zoom for this model on the time window labeled by the black box on panel d1. In this panel, we represent $\left\langle B_\varphi\right\rangle_{\varphi}$ at $r=0.98 R_{\rm top}$ for which we filtered signals with periodicities larger than 5 years to make the short cycle appear more clearly. Finally, panels g to i represent model sO1, which is non-cyclic and close to the anti-solar differential rotation transition.}
  \label{fig:3repcases}
\end{figure*}

\begin{deluxetable*}{lccccccccc}[h]
  \tablecaption{Input parameters and global properties of the models\label{ta:Params}}
  \tablecolumns{9}
  \tablecomments{The subscripts $b$ and $t$ indicate that the quantities where derived from averages in longitude, latitude, and over radial bands $r \in [0.75 R_{\rm top}, 0.8 R_{\rm top}]$ and $r \in [0.95 R_{\rm top}, 0.97 R_{\rm top}]$.} 
  \tabletypesize{\scriptsize}
  \tablehead{
    \colhead{Case} &
    \colhead{$\Omega_\star/\Omega_\odot$} &
    \colhead{$\Delta S$ [$\frac{\rm J}{\rm kg K}$]} &
    \colhead{$N_\rho$} &
    \colhead{$\mathrm{Ro}_b$} &
    \colhead{$L_{b}/L_\odot$} &
    \colhead{Cycle} & 
    \colhead{$\mathrm{Ro}_t$} &
    \colhead{$L_{t}/L_\odot$} & 
    \colhead{Short Cycle}
  }
  \startdata 
sO1 & $0.33$ & $1.00$ & $3.22$ &  $1.39 \pm 0.12$ &   $0.29 \pm 0.01$
&      No &  $4.86 \pm 0.69$ &   $1.17$ &         No \\ 
O1  & $0.55$ & $1.00$ & $3.22$ &  $0.81 \pm 0.09$ &   $0.31 \pm 0.01$ &      Yes &  $2.73 \pm 0.38$ &   $1.21$ &         No \\
O2  & $0.69$ & $1.00$ & $3.22$ &  $0.63 \pm 0.06$ &   $0.33 \pm 0.01$ &      Yes &  $2.12 \pm 0.30$ &   $1.26$ &         No \\
O3  & $0.83$ & $1.00$ & $3.22$ &  $0.50 \pm 0.05$ &   $0.34 \pm 0.01$ &      Yes &  $1.69 \pm 0.23$ &   $1.28$ &         No \\
O4  & $1.00$ & $1.00$ & $3.22$ &  $0.39 \pm 0.05$ &   $0.31 \pm 0.01$ &      Yes &  $1.26 \pm 0.16$ &   $1.09$ &         No \\
O5  & $1.10$ & $1.00$ & $3.22$ &  $0.34 \pm 0.05$ &   $0.30 \pm 0.01$
&      Yes &  $1.11 \pm 0.14$ &   $1.07$ &         No  \\
O6  & $1.25$ & $1.00$ & $3.22$ &  $0.29 \pm 0.05$ &   $0.27 \pm 0.01$ &      Yes &  $0.93 \pm 0.10$ &   $0.92$ &        Yes\\
O7  & $1.66$ & $1.00$ & $3.22$ &  $0.21 \pm 0.04$ &   $0.27 \pm 0.01$ &      No &  $0.64 \pm 0.07$ &   $0.86$ &         Yes \\
O8  & $3.00$ & $1.00$ & $3.22$ &  $0.10 \pm 0.02$ &   $0.28 \pm 0.02$ &      No &  $0.32 \pm 0.03$ &   $0.78$ &         Yes \\
O9  & $5.52$ & $1.00$ & $3.22$ &  $0.04 \pm 0.01$ &   $0.20 \pm 0.02$ &      No &  $0.12 \pm 0.02$ &   $0.39$ &         Yes\vspace*{0.1cm}  \\
S1  & $1.10$ & $0.80$ & $3.22$ &  $0.31 \pm 0.05$ &   $0.19 \pm 0.01$ &      Yes &  $0.92 \pm 0.10$ &   $0.60$ &        Yes \\
S2a & $0.55$ & $1.50$ & $3.22$ &  $1.01 \pm 0.07$ &   $0.55 \pm 0.01$ &      Yes &  $3.88 \pm 0.52$ &   $2.60$ &         No \\
S2b & $1.10$ & $1.50$ & $3.22$ &  $0.44 \pm 0.04$ &   $0.56 \pm 0.02$ &      Yes &  $1.54 \pm 0.20$ &   $2.37$ &         No\vspace*{0.1cm}  \\
R0  & $1.10$ & $1.00$ & $1.00$ &  $0.87 \pm 0.13$ &  $11.34 \pm 2.82$ &      No &  $0.82 \pm 0.04$ &  $17.46$ &         No  \\
R1  & $1.10$ & $1.00$ & $2.00$ &  $0.59 \pm 0.05$ &   $1.89 \pm 0.22$ &      Yes &  $1.08 \pm 0.09$ &   $4.58$ &         No \\
R2  & $1.10$ & $1.00$ & $3.50$ &  $0.31 \pm 0.05$ &   $0.19 \pm 0.01$ &      Yes &  $1.05 \pm 0.13$ &   $0.66$ &         No  \\
R3  & $1.10$ & $1.00$ & $4.00$ &  $0.25 \pm 0.05$ &   $0.10 \pm 0.01$ &      No &  $0.94 \pm 0.13$ &   $0.32$ &         Yes 
  \enddata
\end{deluxetable*}

The various dynamo states realized in our set of simulations occur in particular ranges of the adimensional fluid Rossby number. It can be defined as
\begin{equation}
\mathrm{Ro} = \frac{\left\langle \left| \boldsymbol{\nabla}\times {\bf u}\right| \right\rangle_{\theta, \varphi, t}}{2 \Omega_\star} \label{eq:Rossby}\, ,
\end{equation}
where $\left\langle\right\rangle_{\theta,\varphi,t}$ stands for the average over time, latitude and longitude.
The Rossby number quantifies the importance of rotation in the
dynamics of the system. We define here two Rossby numbers averaged
over the [$0.75 R_{\rm top}$,$0.8 R_{\rm top}$] domain
($\mathrm{Ro}_b$, where the deep-seated magnetic field lies) and over
the [$0.95 R_{\rm top}$,$0.97 R_{\rm top}$] domain ($\mathrm{Ro}_t$,
where the short cycle is observed in the simulations developing
it). Both Rossby numbers are reported in Table \ref{ta:Params}, and
note that the averaging intervals were chosen to safely exclude the
radial boundaries. Our series of simulations covers a range of Rossby
numbers $\mathrm{Ro}_b$ between $0.04$ and $1.39$, and $\mathrm{Ro}_t$
between $0.12$ and $4.86$ (see Table \ref{ta:Params}). The error-bars
in Rossby numbers are taken from the standard deviation of
$\mathrm{Ro}$ over the radial averaging interval. For the sake of
completeness, we also report in Table \ref{ta:Params} the convective
luminosities defined as \reff{$L(r) = 4\pi r^2 \bar{\rho} c_P
  \left\langle u_r T\right\rangle_{\theta, \varphi,t}$} at these two
locations as well \reff{(with $T$ representing the temperature fluctuations
with respect to the ambient state)}. 

We summarize in Fig. \ref{fig:SummaryPlot} our ensemble of dynamo states, where the ratio between the time-averaged kinetic energy contained in the differential rotation ('DRKE', see Appendix \ref{sec:DefEnergies}) and the \reff{total kinetic energy ('KE')} is shown as the function of the Rossby number $\mathrm{Ro}_b$. The color of the labels indicates the dynamo state each model achieves, which are also reported at the top of the figure. We immediately see that three regimes of Rossby numbers exist in our set of simulations. When $\mathrm{Ro}_b$ is below 0.3, a short magnetic cycle is observed near the surface and close to the equator (blue symbols). For $0.25 \lesssim \mathrm{Ro}_b \lesssim  1$, a deep-seated magnetic cycle is realized (red symbols). Finally, as soon as $\mathrm{Ro}_b > 1$, the magnetic cycles disappear and stable wreaths of magnetic field are observed at the base of the convective envelope (black symbols). Note that two of the simulations in our set (namely 06 and S1) possess both a deep-seated cycle and a short cycle at the same time, which was also reported in the millenium EULAG simulation \citep[see, e.g.][and references therein]{Beaudoin:2016kt}. \modifASt{The short cycle close to the surface presents a poleward migration (panel d2 in Fig. \ref{fig:3repcases}), which is consistent here with the Parker-Yoshimura \citep{Parker:1955km,Yoshimura:1975bu} rule of a dynamo wave propagating along the near-cylindrical iso-contours of differential rotation. The deep-seated cycle does not follow such a rule, and originates from the non-linear feedback of the Lorentz force on the differential rotation itself, as will be made clear in \S\ref{sec:prot-cycl-solut} and \S\ref{sec:dynamo-action}.} 

\modifPC{It has been noted before} that cyclic dynamos are observed over a limited range of Rossby numbers,
starting with the pioneering work of \citet{Gilman:1983dx} (see, \textit{e.g.}, his figure 31). We observe here a similar trend, where the low-Rossby number transition to cyclic solutions corresponds to a threshold value of DRKE/KE of 0.6 (around $\mathrm{Ro}_b \sim 0.25$). The high-Rossby number transition is found at a lower threshold (DRKE/KE $\sim$ 0.1 to 0.2), and is found to occur when the differential rotation switches from being solar-like (fast equator, slow poles, see panel e in Fig. \ref{fig:3repcases}) to anti-solar (slow equator, fast higher latitudes, \textit{e.g.} panel h in Fig. \ref{fig:3repcases}). We discuss in more details this transition in \S\ref{sec:large-rossby-number}. We now briefly characterize the energetics of our simulation set before detailing the magnetic cycles themselves in subsequent sections.

\begin{figure}[htb]
  \centering
  \includegraphics[width=\linewidth]{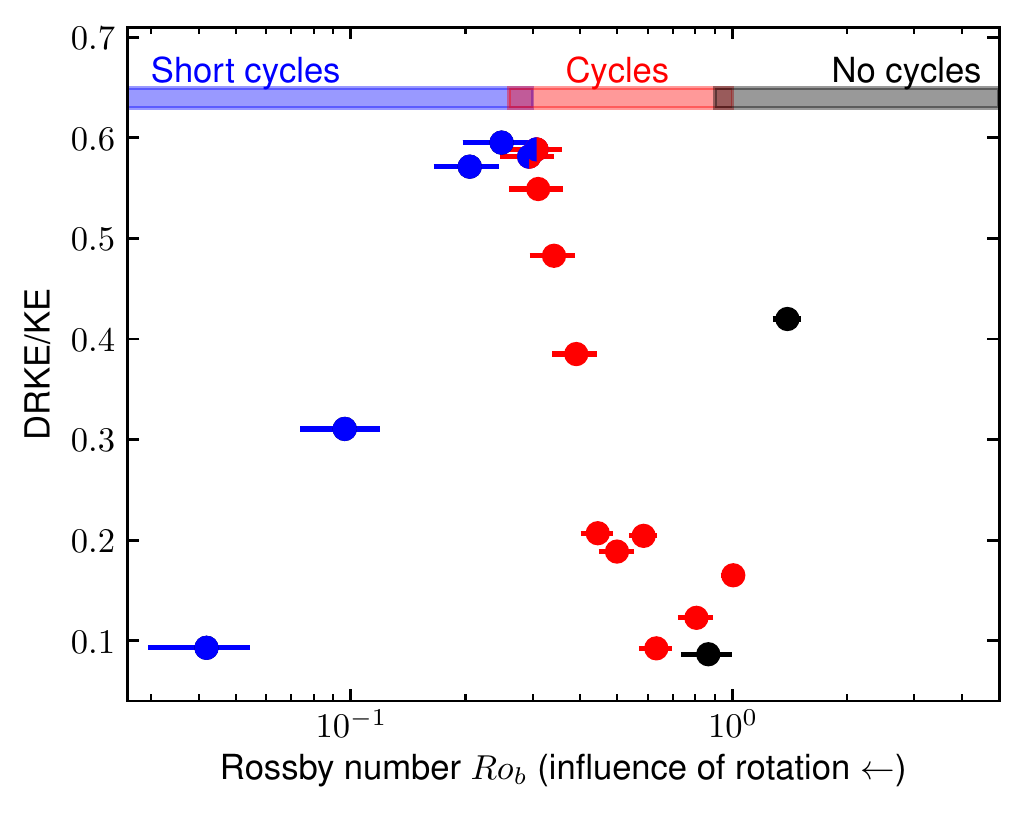}
  \caption{Ratio of the differential rotation kinetic energy to the
    total kinetic energy \refff{(third column in Table
      \ref{ta:energetics})} as a function of the Rossby number
    $\mathrm{Ro}_b$ for all the models in our set of simulations
    \reff{The errorbars are given in Table \ref{ta:Params}.} The color bars at the top label the Rossby number regimes where we observe a short cycle (blue), deep-seated cycle (red) and no cycle at all (black).}
  \label{fig:SummaryPlot}
\end{figure}


\subsection{Energetics in the simulation set}
\label{sec:energetics-sim-set}

The simulations in our set strongly vary in their kinetic and magnetic energy properties, which we list in Table \ref{ta:energetics} (see Appendix \ref{sec:DefEnergies} for the definitions of the various energies). We find that the differential rotation (DRKE, third column) accounts for a significant part of the kinetic energy in most cases, between $10\%$ and $60\%$ of the total energy (KE, second column). The convective kinetic energy (CKE, fourth column) accounts for more than $40\%$ of the total kinetic energy in all cases.

The majority of our simulations develop a dynamo in sub-equipartition, with magnetic energy (ME, fifth column) of the order of $10\%$ of the total kinetic energy. The two simulations (O8 and O9) at the higher rotation rates, for which the specifics of the dynamo solution
are discussed in \S\ref{sec:low-rossby-number}, achieve a super-equipartition regime up to ME$\sim$ 5 KE (magnetostrophic state, see \citealt{Brun:2017dy}). In all other cases, $30\%$ to $50\%$ of the total magnetic energy is distributed roughly equally over the large-scale axisymmetric toroidal (TME, sixth column) and poloidal (PME, seventh column) fields. The fluctuating magnetic energy spectrum peaks at mid-to-small scale (typically around the spherical harmonics degree $l=30$). As a result, the large-scale field is dominated by the axisymmetric dipole and quadrupole, while the small-scale field is dominated by non-axisymmetric modes.


\begin{deluxetable*}{lccccccc}
  \tablecaption{Energetics of the models\label{ta:energetics}}
  \tablecolumns{8}
  \tabletypesize{\scriptsize}
  \tablehead{
    \colhead{Case} &
    \colhead{KE} &
    \colhead{DRKE} & 
    \colhead{CKE} &
    \colhead{ME} & 
    \colhead{TME} & 
    \colhead{PME} &
    \colhead{FME} 
\\
    \colhead{} &
    \colhead{[$10^{31}$ J]} &
    \colhead{[$\%$ KE]} & 
    \colhead{[$\%$ KE]} &
    \colhead{[$\%$ KE]} & 
    \colhead{[$\%$ ME]} & 
    \colhead{[$\%$ ME]} &
    \colhead{[$\%$ ME]} 
  }
  \startdata 
\input{EnergyTable}
  \enddata
\end{deluxetable*}

}




\section{A representative cyclic solution}
\label{sec:prot-cycl-solut}

\modifAS{We now focus on a reference solution displaying features found, to some extent,
in all the cyclic simulations presented above.}
We opted to show simulation O5, which is rotating slightly faster than the Sun
and has a solar-like convective luminosity at its top (as shown in Table \ref{ta:Params}). 
The magnetic energy amounts to a non-negligible fraction
of the kinetic energy ($\sim$10\%), the \reff{later} being dominated by the fluctuating
components ('CKE').

\begin{figure}[htb]
  \centering
  \includegraphics[width=\linewidth]{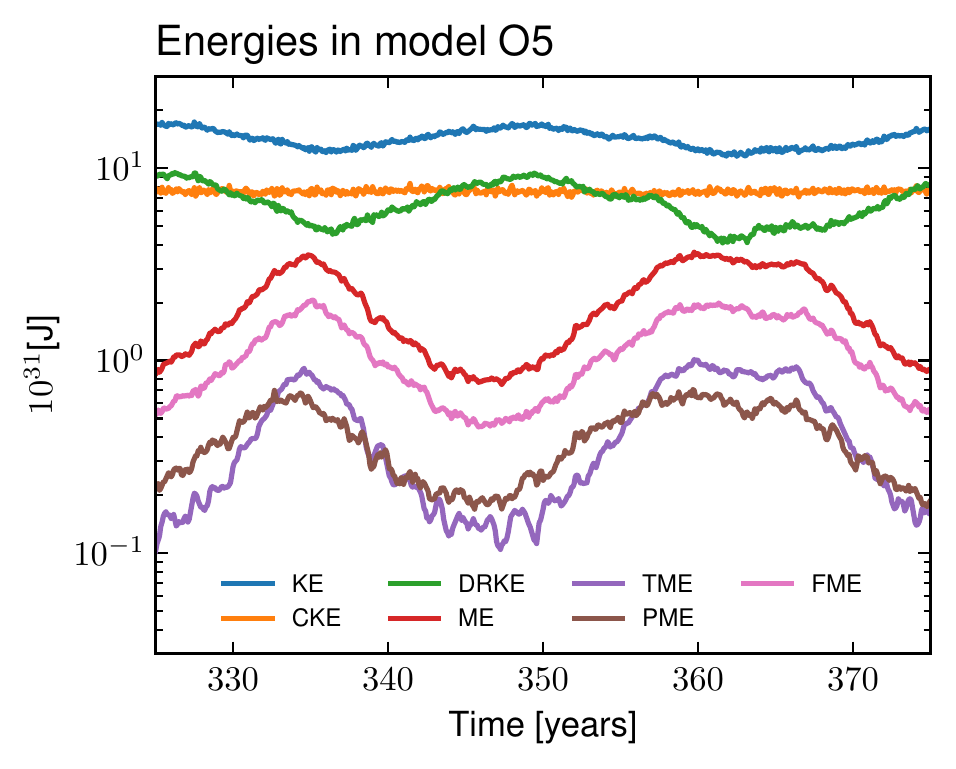}
  \caption{Temporal series of energies taken over a magnetic cycle and integrated over
  the whole domain in model O5. Blue and red are respectively the total kinetic and magnetic energies.
  Orange is the kinetic energy related to convection, and green is the differential
  rotation energy. Purple is the magnetic energy in the mean toroidal field, while brown and pink are respectively
  energies in the mean poloidal field and in the non-axisymmetric components of the magnetic field.}
  \label{fig:Energies}
\end{figure}

Fig.~\ref{fig:Energies} shows time series of the energy contributions in simulation O5
over a limited time interval spanning two magnetic reversals. Almost all energies are temporally modulated on both long (of the order of 20 years) and short (less than a year) timescales. The total magnetic energy (red) shows a cyclic modulation anti-correlated with the kinetic energy of the differential rotation (green), 
indicating a coupling between the large-scale flows and fields through the Lorentz force
(as further investigated in \S\ref{sec:dynamo-action}). Note that the energy variation in both of them is of comparable amplitude, which suggests that the energy transfer between the two plays a significant role in the dynamo action. 
The energy of the turbulent, convective motions (orange) appears to be insensitive to these decadal variations, similarly to previously published EULAG and ASH global MHD simulations presenting magnetic cycles \citep[\textit{e.g.}][]{Brun:2005fv,Racine:2011gh,Augustson:2015er}. The toroidal (purple) and fluctuating (pink) magnetic energies are found to vary in phase with the total energy. The poloidal magnetic energy (brown) is also essentially varying in phase with the total magnetic energy, albeit small time-lags are observed for some reversal (around $t \sim 334$ yr in Fig. \ref{fig:Energies}) which can be traced to hemispheric decorrelations.

\begin{figure*}[tb]
  \centering
  \includegraphics[width=\linewidth]{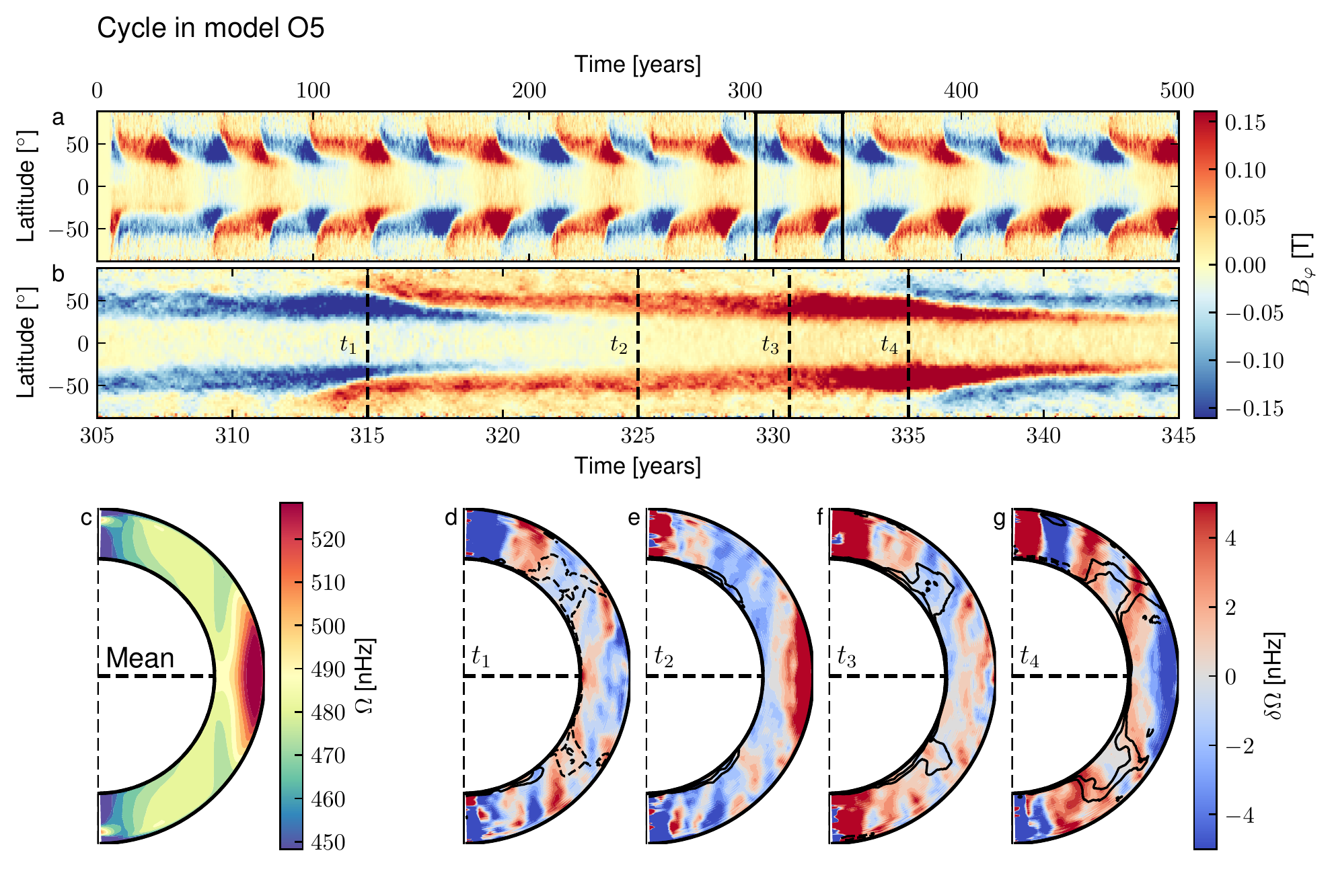}
  \caption{Combination of diagrams showing the toroidal magnetic field and its impact
  on the differential rotation profile in model O5. The top two panels display latitude-time representations
  of the toroidal magnetic field, near the base of the convection zone 
  ($r = 0.75\,R_{\mathrm{top}}$). Panel a shows the field on the full length of the
  simulation, while panel b spans the time interval indicated by the
  black box in the top diagram. The bottom diagrams
  show the mean differential rotation profile (panel c) and the departure from the mean differential rotation profile (panels d to g) at four different times (shown in dashed lines in the panel just
  above them). Solid (dashed) black lines in the bottom left diagrams denote the 0.1 T and 0.2 T isocontours of positive (negative) toroidal magnetic field.}
  \label{fig:BphiDr}
\end{figure*}

An example of a magnetic cycle is shown in Fig.~\ref{fig:BphiDr} for model O5, along
with the mean differential rotation profile (panel c) and its variations during one magnetic reversal (panels d to g).
The two top panels show the longitudinally averaged toroidal magnetic fields in latitude-time representations, taken in the lower half of the convection zone
($r = 0.75\, R_{\mathrm{top}}$). Panel a covers the full length of the
simulation, with a rectangular box delimiting the smaller time interval displayed in panel b, covering one magnetic cycle.
Some solar-like features
in the large-scale magnetic field are apparent here, such as regular inversions of polarities,
a slight propagation towards the equator during a half-cycle, and magnetic intensities comparable to the ones expected at the base of the solar convection zone. However, the field is quadrupolar, not entirely well synchronized between the hemispheres and located at mid- to high latitudes (some cases exhibit a beating between dipolar and quadrupolar symmetry, see \S\ref{sec:dynamo-families}).

\modifAS{This magnetic cycle
is the source of the signatures seen in Fig.~\ref{fig:Energies} in both
magnetic and differential rotation energies. Panel c of Fig.~\ref{fig:BphiDr} shows the time-averaged, solar-like differential rotation profile 
$\left\langle \Omega \right\rangle_{t}$ of model O5 with a fast equator and slow poles, the rotation being
defined as $\Omega = \Omega_\star + \left\langle u_\varphi \right\rangle_{\varphi}/\left(r\sin\theta\right)$.
At the bottom right, the departure from the mean profile
$\delta\Omega = \Omega - \left\langle \Omega \right\rangle_t$ is shown at
four time steps corresponding to the dashed lines in panel b. These time steps span different phases of the magnetic cycle: from a
maximum ($t_1$, panel d) to a minimum ($t_2$ and $t_3$; panels e and f), then to a maximum again ($t_4$, panel g).
Fast (slow) angular velocities compared to the mean are represented in red (blue).
Isocontours of positive (negative) toroidal magnetic field are overlaid in black solid (dashed) lines. 
During the phases of maximum ($t_1$, $t_4$; panels d and g), strong accelerations (red) of the differential rotation are observed (see also \S\ref{sec:dynamo-action}) on both sides of the magnetic structures (black lines). During the miminum of the cycle ($t_2$, panel e), the overall differential rotation is reinforced (acceleration at the equator, deceleration at mid-latitudes) as the simulation is closer to a purely hydrodynamic state \citep{Beaudoin:2018tl}. This phenomenon is at the heart of the dynamo action taking place in these simulations, which we now turn to.}

\section{Dynamo action}
\label{sec:dynamo-action}

The dynamo action sustaining the cycling magnetic field described in \S\ref{sec:prot-cycl-solut} requires further investigation. We first examine the polarity inversion mechanism (\S\ref{sec:polar-invers-mech}) and then characterize the symmetry properties of the dynamo solution (\S\ref{sec:dynamo-families}). 

\begin{figure*}[htb]
  \centering
  \includegraphics[width=\linewidth]{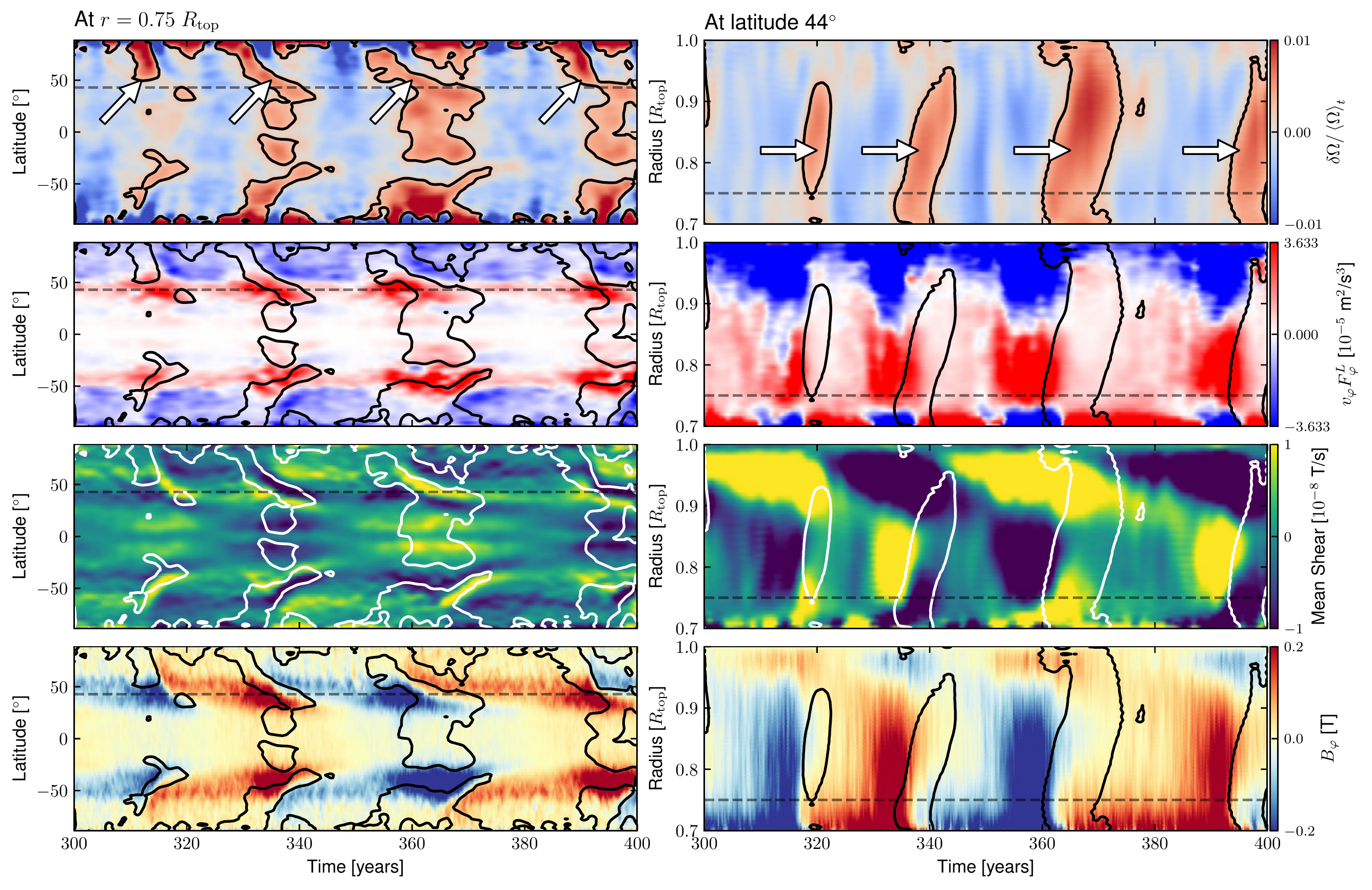}
  \caption{Dynamo reversal mechanism in model O5. The left panels represent time-latitude diagrams at $r=0.75\,R_{\rm top}$, right panels radius-latitude diagrams at $\theta = 44^\circ$ for two cycle periods. The black dashed lines label the latitude and radii at which the panels of the other column are shown. The first row shows the differential rotation perturbation $\delta\Omega$ (see text) from the time-averaged differential rotation, with white arrow labeling the acceleration phases. The black contours trace the $0.25\%$ variation level and are overlayed in black in the second and fourth rows, and in white in the third row. The second row shows the energy transfer from magnetic energy to azimuthal kinetic energy, red denoting a positive transfer from magnetic to kinetic. The third row shows the mean shear of the poloidal field by the differential rotation (see text), yellow being positive and blue negative. Finally, the fourth row show the mean azimuthal field with blue denoting negative $B_\varphi$ and red positive $B_\varphi$.}
  \label{fig:DynamoReversal}
\end{figure*}
\subsection{Polarity inversion mechanism}
\label{sec:polar-invers-mech}

In all simulations presenting a deep-seated primary cycle, we observe temporal modulations of the mean differential rotation correlated with the magnetic cycle (see Fig. \ref{fig:BphiDr}). We propose here that these modulations play a major role in allowing the magnetic cycle to operate, and in setting its period. Let us first characterize in more details this modulation, shown in the top panels of Fig. \ref{fig:DynamoReversal} for model O5 (the left panels represent time-latitude diagrams at $r=0.75 R_{\rm top}$, right panel display radius-latitude diagrams at latitude $44^\circ$). The modulation of the differential rotation $\delta\Omega/\left\langle\Omega\right\rangle_t$ amounts to $\sim$1\% of the total differential rotation in case O5, which corresponds to $\sim$50\% when the rotating frame $\Omega_\star$ is subtracted. Systematic accelerations (in red, contoured for 0.25 \% in black, and highlighted by white arrows to guide the eye) are correlated with the magnetic cycle, propagating towards the equator (left panel) and towards the surface (right panel). 

We first demonstrate that these modulations originate from the
feedback of the magnetic field on the differential rotation. In the
second row of Fig.~\ref{fig:DynamoReversal}, we show the energy
transfer due to the Lorentz force towards the \reff{toroidal}
kinetic energy \reff{(last term in Eq. \ref{eq:LBRET_u}, see also Eqs B.12 and B.13 in \citealt{Nelson:2013fa})}. A positive (red) transfer means that energy is effectively transferred from magnetic to kinetic, and negative (blue) indicates the opposite. We observe that locally, at mid-latitude and in the lower part of the convection zone, energy is transferred from magnetic to kinetic energy. The energy transfer furthermore peaks just before the cyclic acceleration in the upper panel (the contours of $\delta\Omega$ have been overlaid here for clarity). The positive transfer that is sustained over roughly 100 solar days is indeed able to induce a flow of magnitude of $\sim 10$ m/s, which is on par with the acceleration displayed in the top panels. As a result, the acceleration phases observed in our simulations are indeed of magnetic origin. Note that these modulations differ from the one observed in the millenium EULAG simulation \citep{Beaudoin:2016kt}, which were shown to originate indirectly from the magnetically-induced meridional circulation modulations \citep{Passos:2012ht}. Analogous modulations of $\Omega$ were also reported in the simulations of \citet{Augustson:2015er} \reff{and} were shown to be at the origin of the equatorward migration of their magnetic structures. Here the equatorial migration is very weak, but the direct feedback of the magnetic field on the differential rotation is stronger (see energies in Fig. \ref{fig:Energies}).

However, are these modulations affecting dynamo action? We display the
equivalent of the dynamo Omega-effect in our 3D non-linear simulations
in the third row of Fig. \ref{fig:DynamoReversal}. The mean shearing
of the magnetic field (or $\Omega$-effect),
$\left. \left(\left\langle{\bf B}\right\rangle_p
    \cdot\boldsymbol{\nabla}\right)\left\langle{\bf
      u}\right\rangle_\varphi \right|_\varphi$, is represented in
time-latitude and time-radius diagrams, with the same contours of
$\delta\Omega$ overlaid in white. We observe that immediately during
the acceleration phases of the differential rotation, the mean shear
severely drops in amplitude, and \reff{ultimately} locally changes
sign (going from yellow to blue or blue to yellow) due to a change of
sign in the latitudinal gradient of the differential rotation (see,
\textit{e.g.}, Fig. S8 in \citealt{Strugarek:2017go}). \reff{The azimuthal
field consequently ceases to be sustained (last row in
Fig. \ref{fig:DynamoReversal}), and as a result the poloidal field (not shown here, see \citealt{Strugarek:2017go}) also loses its source and rapidly decreases following the azimuthal field decrease.}

As the magnetic field wanes, the acceleration of the differential rotation fades away as the energy transfer mediated by the Lorentz force severely weakens (see second row of panels). During this phase, the mean shear (third panel) is still of the opposite polarity and thus a weak azimuthal field of opposite polarity is generated. This new seed azimuthal field is strong enough so that a poloidal field (also of opposite polarity) is slowly generated as well. The differential rotation finally recovers its ground state (blue phases in the first panel), and the dynamo process is able to amplify again both components of the field, albeit with opposite polarity compared to the previous configuration. 
\modifPC{Eventually,}
the field becomes large enough such that an efficient energy transfer again perturbs locally the differential rotation, which triggers a new reversal. This mechanism repeats over and over, and similarly so in all the simulations of our set which generate a deep-seated magnetic cycle.

\subsection{Dynamo families}
\label{sec:dynamo-families}

\begin{figure}[htb]
  \centering
  \includegraphics[width=\linewidth]{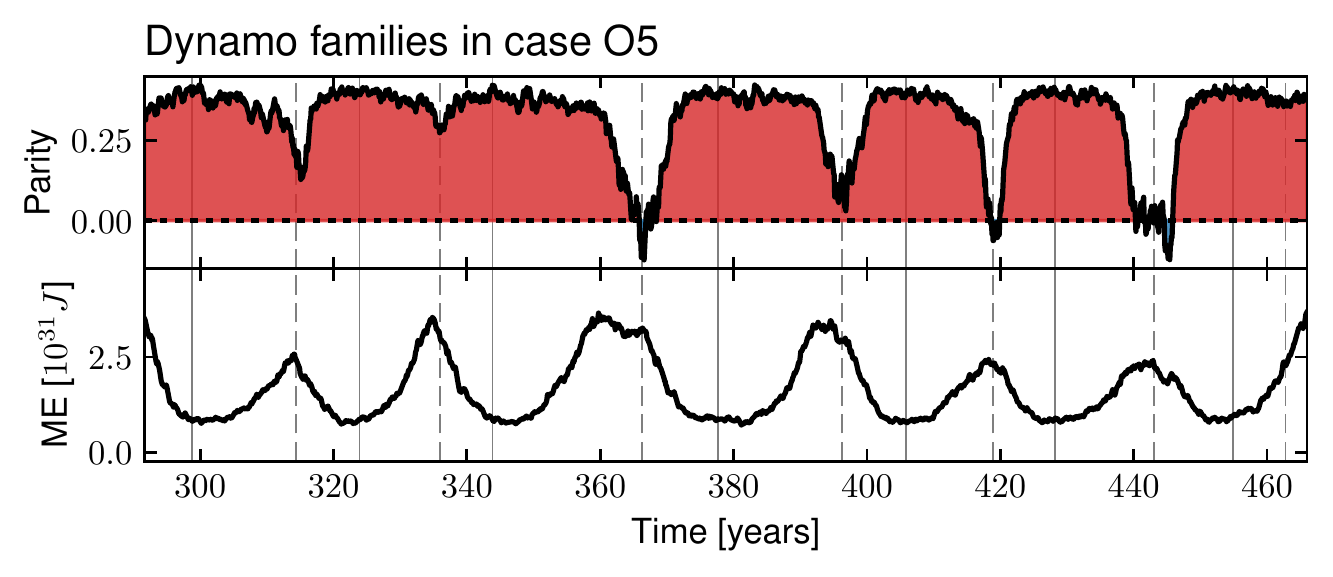}
  \includegraphics[width=\linewidth]{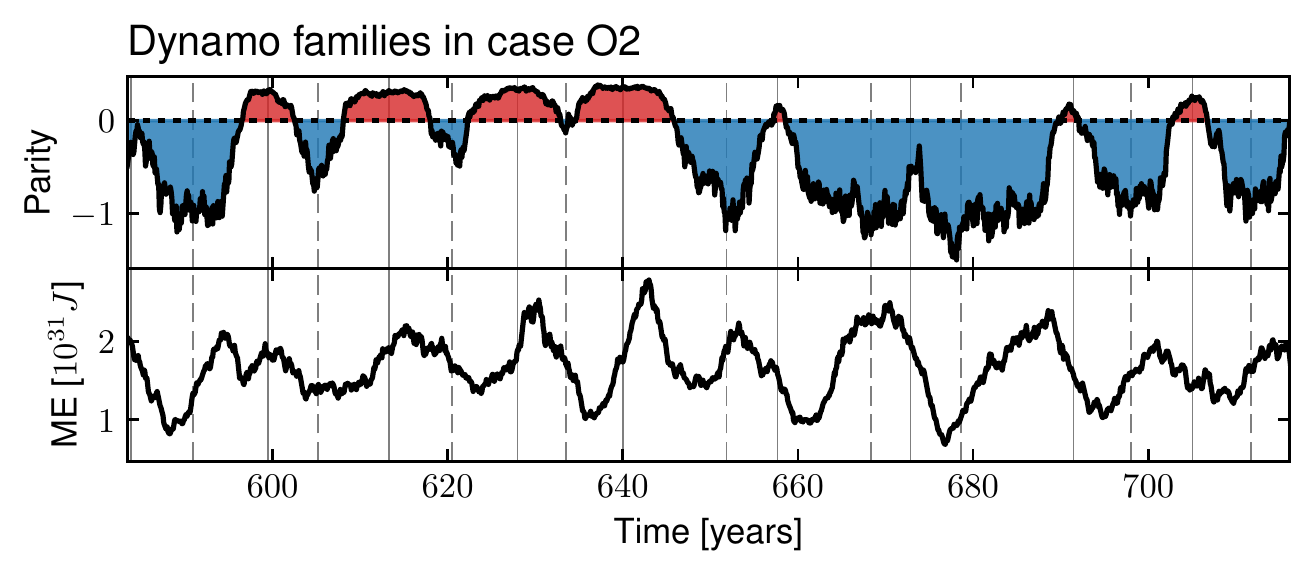}
  \caption{Dynamo families (top panels) and magnetic energy (bottom panels) for cases O5 (top) and O2 (bottom). The parity $P$ (Eq. \ref{eq:parity}) is colored in red when dominated by a quadrupolar symmetry ($P>0$) and in blue when dominated by a dipolar symmetry ($P<0$). Vertical plain and dashed line respectively label phases where quadrupolar and dipolar symmetries are maximized.}
  \label{fig:DynamoFamilies}
\end{figure}

Most of the dynamo states achieved in this series of simulations present a quadrupolar symmetry, with \textit{e.g.} an azimuthal field symmetric with respect to the equator (see Fig. \ref{fig:BphiDr}). This appears to be contrary to the solar dynamo, which is dominated by a dipolar symmetry. We characterize the dynamo solutions by calculating the energy repartition between dipolar ('anti-symmetric') and quadrupolar ('symmetric') families \citep{Roberts:1972wu,Gubbins:1993ii,McFadden:1991bw}. Projecting a vector on the vectorial spherical harmonics basis \citep{Rieutord:1987go,Mathis:2005kz,Strugarek:2013kt} 
\begin{equation}
\label{eq:RSTBasis}
	{\bf X} = \sum_{l,m} A^l_m \boldsymbol{\mathcal{R}}^m_l +  B^l_m \boldsymbol{\mathcal{S}}^m_l +  C^l_m \boldsymbol{\mathcal{T}}^m_l\, ,
\end{equation}
we can separate its quadrupolar and dipolar components as
\begin{eqnarray}
\label{eq:quadcomp}
	{\bf X}_Q &=&  \cdots + A^{m}_{m}\boldsymbol{\mathcal{R}}^{m}_{m} +
  B^{m}_m\boldsymbol{\mathcal{S}}^{m}_{m} +
  C^{m+1}_m\boldsymbol{\mathcal{T}}^{m}_{m+1} 
  + \dotsc \, ,
\\
	{\bf X}_D &=&  \cdots + A^{m+1}_{m}\boldsymbol{\mathcal{R}}^{m}_{m+1} +
  B^{m+1}_m\boldsymbol{\mathcal{S}}^{m}_{m+1} +
  C^{m}_m\boldsymbol{\mathcal{T}}^{m}_{m}
  + \dotsc \, .
\label{eq:dipcomp}
\end{eqnarray}
Based on this decomposition, we further define the parity of the magnetic field \citep{Tobias:1997vc} with the ratio of
magnetic energies $E_Q$ and $E_D$
\begin{equation}
\label{eq:parity}
	P = \frac{E_Q-E_D}{E_Q+E_D}\, .
\end{equation}
The parity P is equal to $+1$ when the field is mostly quadrupolar, and $-1$ when it is mostly dipolar. The parity of two representative simulations (O5 and O2) is shown in Fig. \ref{fig:DynamoFamilies} at the top of the domain, along with the total magnetic energy ME (Eq. \ref{eq:ME}).

Case O5 (Fig. \ref{fig:BphiDr}) exhibits a dynamo solution dominated by the quadrupolar symmetry at all times. When the total magnetic energy peaks, the parity falls down to zero in this case (dipoles and quadrupoles have the same energy). This is somehow opposite to what is observed in the Sun, with a dipolar symmetry dominating during cycle minimum and a quadrupolar symmetry dominating at cycle maximum \citep{DeRosa:2012gx}. Case O2 presents an interesting beating between the two families over a time-scale larger than the magnetic cycle itself. Between $t = 590$ yr and $t = 650$ yr, quadrupolar symmetry dominates the field, as in model O5. Nevertheless, model O2 behaves more like the Sun during this period, with the dipolar symmetry being most prominent during cycle minimum (see, e.g., \citealt{DeRosa:2012gx}). After $t=650$ yr, the polarity trend is reversed and the field is dominated by the dipolar symmetry. 

Co-existing modulation of magnetic parity and large-scale flows
is a well-known property of nonlinear dynamos, which has been 
explored at length using dynamical system approaches and geometrically
simplified mean-field and mean-field-like dynamo models
(e.g., \citealt{Beer:1998ct,Knobloch:1998ev,Weiss:2000ea}, and references therein) and to some extent also recently in non-linear three dimensional simulations \citep{Augustson:2015er,Raynaud:2016hc}. The development of joint modulations
of parity and large-scale flow in our simulations is consistent with
the magnetic polarity reversal scenario proposed in what follows,
under which energy exchange between magnetic and kinetic energy reservoirs
is a key ingredient in the evolution of the magnetic cycle.

\section{Trends in the cycle periods and transitions in dynamo states}
\label{sec:trends-scaling-laws}

\subsection{Characterization of the cycle periods}
\label{sec:char-cyc-peri}

In order to characterize the cycle period, we need a robust methodology to extract the most significant period out of a potentially multi-periodic signal. One such method was recently successfully applied to similar simulations by \citet{Kapyla:2016hg} based on the so-called empirical-mode decomposition (EMD) method (we refer the reader to Appendix \ref{sec:emd} and \citealt{Kapyla:2016hg} for a more in depth discussion about this method). We based our analysis on the \textit{libeemd} library \citep{Luukko:2015jy} to make use of the complete variant ensemble empirical mode decomposition (CEEMDAN) method. This method automatically decomposes a signal on a series of intrinsic mode functions, leveraging the addition of noise to better separate different frequencies in the spectra.

\begin{deluxetable}{lcccccc}
  \tablecaption{Cycles \label{ta:cycles}}
  \tablecolumns{7}
  \tablecomments{Period of the magnetic cycles detected near the bottom ($R_b=0.72\,R_{\rm top}$) and top ($R_t=0.98\,R_{\rm top}$) of the convective envelope with the CEEMDAN analysis. The details of the method are given in Appendix \ref{sec:emd}.}
  \tabletypesize{\scriptsize}
  \tablehead{
    \colhead{Case}  &
    & 
    \colhead{$R_b$} &
    &
    &
   \colhead{$R_t$}
  &
\\
   &
   \colhead{P [yr]} &
   \colhead{E} &
   \colhead{R} &
   \colhead{P [yr]} &
   \colhead{E} &
   \colhead{R} 
  }
  \startdata 
\input{Cycles.tex}
  \enddata
\end{deluxetable}

Based on our analysis of the cases presented in the present study, we perform the CEEMDAN analysis at two representative radii and latitudes, $[R_b=0.72 R_{\rm top},\,\theta=45^\circ]$ and $[R_t=0.98 R_{\rm top},\,\theta = 25^\circ]$ on the longitudinally-averaged azimuthal component of the magnetic field. At the top position, the CEEMDAN analysis is applied to the detrended $\left\langle B_\varphi\right\rangle_\varphi$ (see panel d2 in Fig. \ref{fig:3repcases}). For each position, we identify the most energetic mode as the representative cyclic mode (see Appendix \ref{sec:emd}). The period [P], energy [E] and robustness [R] of the modes are given in Table \ref{ta:cycles}. Their energy is calculated relative to the total energy of all the decomposed modes, and the robustness is estimated by comparing the results with the results of CEEMDAN applied to a white noise time series which has the same length, mean and standard deviation as the analyzed simulation time series (robustness increases as R$\rightarrow$1). The signal of the shorter cycle at $R_t$ is generally weaker than the signal of the primary cycle at $R_b$. As a result, any cycle for which all modes have either (1) a relative energy less than 0.5 for the primary cycle or 0.3 for the short cycle, or (2) a robustness smaller than 0.2, are considered undetected (symbol '$-$' in Table \ref{ta:cycles}). The error in the cycle period is estimated based on the standard deviation of the corresponding empirical mode. We have reproduced this analysis varying slightly the latitude at the chosen depths $(R_b,R_t)$ without finding any significant changes in our results, all detected cycle periods being well located within their error-bars.

We find that, in our sample of models, deep-seated magnetic cycles materialize when $0.25 \lesssim \mathrm{Ro}_b \lesssim 1$, and short cycles when $\mathrm{Ro}_b \lesssim 0.3$ (or equivalently $\mathrm{Ro}_t \lesssim 1$), as shown in Fig. \ref{fig:SummaryPlot}. Only two models in our set manage to sustain simultaneously the two types of cycles (models O6 and S1). When the Rossby number is significantly larger than 1, no cyclic magnetism could be realized in our simulations (we will come back to this point in \S\ref{sec:large-rossby-number}). 

\subsection{Trends of cyclic solutions}
\label{sec:trends-cycl-solut}

\begin{figure}[htb]
  \centering
  \includegraphics[width=\linewidth]{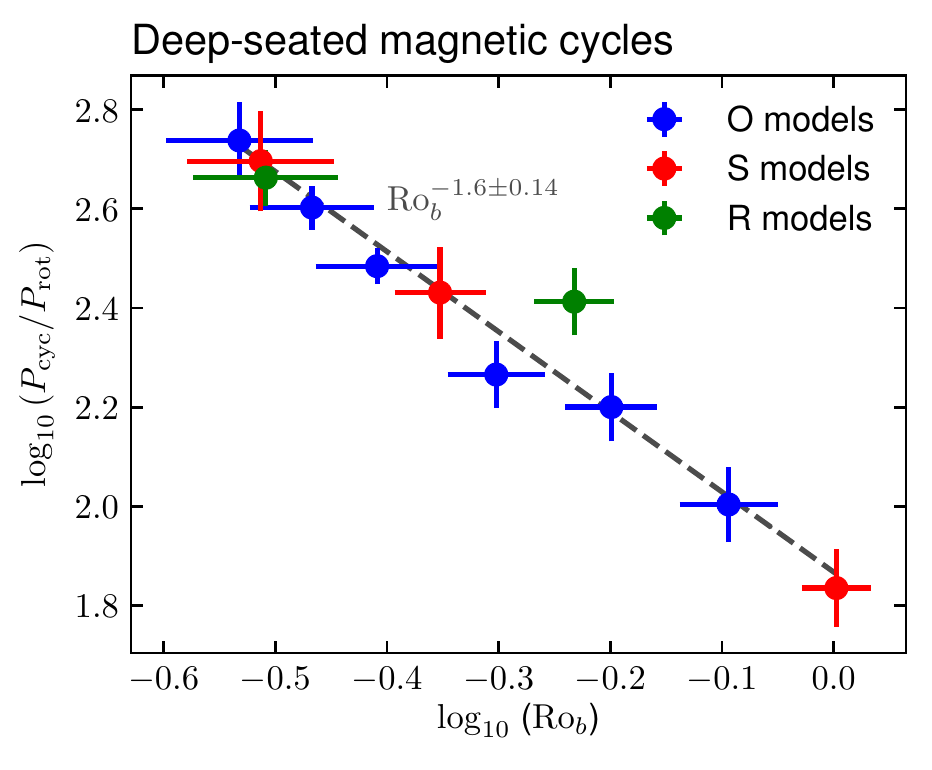}
  \caption{Cycle period normalized to the stellar rotation period as a function of the Rossby number $\mathrm{Ro}_b$ in logarithmic scale. The different models are color-coded as models where the rotation rate was changed ('O' models, in blue), the entropy contrast in the convective envelope was changed ('S' models, in red), and the number of density scale heights was changed ('R' models, in green). Orthogonal distance regression was performed to find the power-law fitted indicated by the gray dashed line, using the error-bars for both the estimated cycle period (Table \ref{ta:cycles}) and Rossby number (Table \ref{ta:Params}).}
  \label{fig:RossbyTrend}
\end{figure}

The period of the deep-seated magnetic cycle decreases with increasing Rossby number (viz. Tables \ref{ta:Params} and \ref{ta:cycles}). Furthermore, the cycle period decreases like a power-law with an exponent of $-1.6 \pm 0.14$ when normalized to the rotation period of the star, as shown in Fig. \ref{fig:RossbyTrend}. This exponent is slightly 
\modifPC{smaller}
than the one derived in \citet{Strugarek:2017go}, where a smaller sample of cyclic models was used. A slope smaller than $-1$ nevertheless remains a robust feature of our non-linear 3D simulations. All models fit relatively well one single power-law dependency with the Rossby number $\Ro_b$. 

\modifAS{Model R1 (green dot in the middle of Fig. \ref{fig:RossbyTrend}) is the most distant point to the power-law fit, suggesting that the density contrast in the convective envelope slightly affects the power-law dependency of the cycle period to the Rossby number found in this work. The strong effect of stratification on the dynamo state was already reported by \citet{Kapyla:2014jr}, who showed that the transition to a cyclic dynamo occurred at higher Rossby numbers as the density stratification was increased. In the present case, additional cyclic solutions with larger density contrasts would be required to assess precisely the effect of this parameter on the power-law fit. However, the density contrast in stellar convective envelopes changes accordingly with their aspect ratio. As a result, both effects ultimately need to be studied jointly to fully characterize how magnetic cycles depend on density contrast, which we leave here for future work.}

\begin{figure}[htb]
  \centering
  \includegraphics[width=\linewidth]{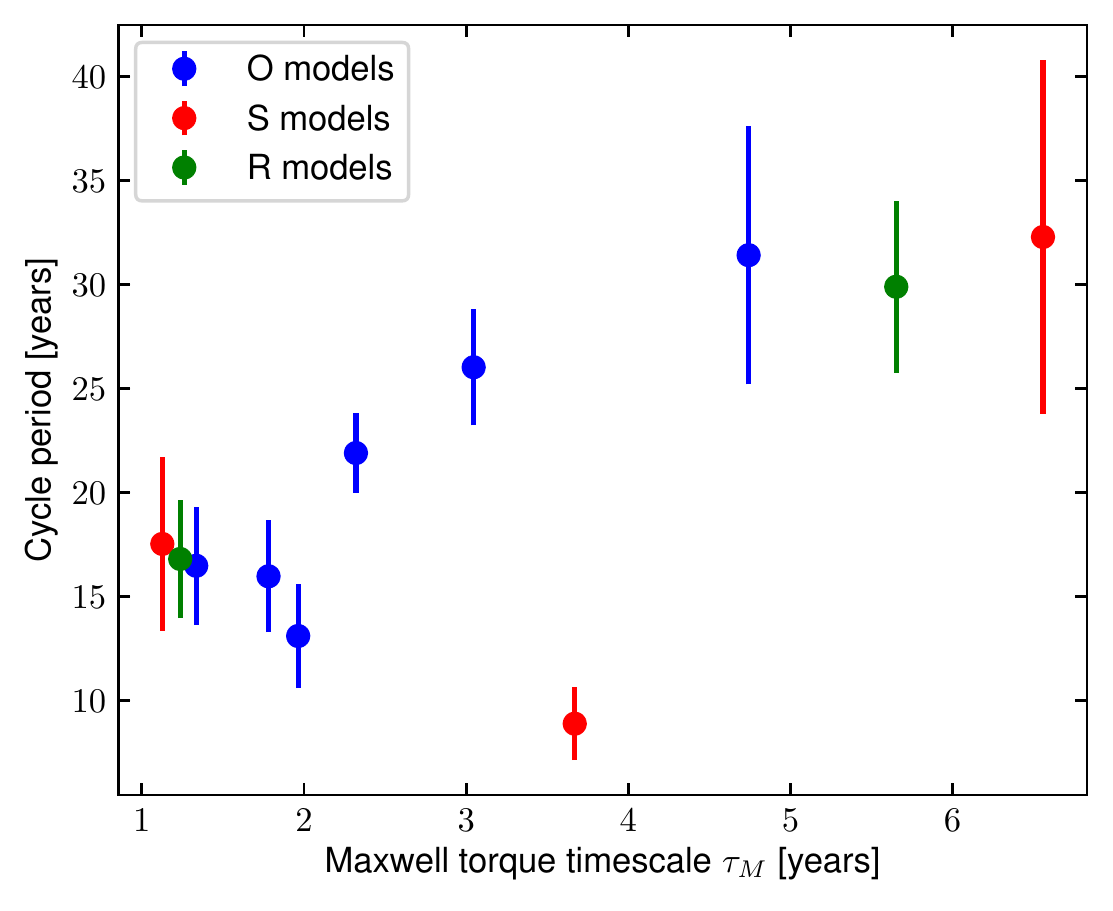}
  \caption{Deep-seated magnetic cycle period as a function of the Maxwell torque timescale $\tau_M$ (Eq \ref{eq:MaxwellTorqueTime}). The models are color-coded as in Fig. \ref{fig:RossbyTrend}.}
  \label{fig:TauM}
\end{figure}

The non-linear character of the dynamo action detailed in \S\ref{sec:dynamo-action} was proposed to be at the origin of the negative slope observed in Fig. \ref{fig:RossbyTrend} by \citet{Strugarek:2017go}. Using this interpretation, the magnetic cycle length is set by the Lorentz force feedback on the large-scale differential rotation: the stronger the feedback is, the shorter the cycle will be. This interpretation can be further tested by calculating the timescale $\tau_M$ associated with the Maxwell torque in our simulations. By equating the time derivative in Eq. \ref{eq:LBRET_u} with the right-hand side large-scale Lorentz force, one may estimate $\tau_M$ by
\begin{equation}
\label{eq:MaxwellTorqueTime}
	\tau_M \approx \sqrt{\frac{\rm DRKE}{{\rm TME}\cdot {\rm PME}}} \cdot d \cdot \sqrt{\frac{\rm \mathcal{V}_{CZ} \hat{\rho}}{2}}\, ,
\end{equation}
where $\mathcal{V}_{\rm CZ}$ is the volume of the convection zone, $d = R_{\rm top}-R_{\rm bot}$ is the depth of the convection zone, and $\hat{\rho}$ is the background density averaged over the convection zone. We display in Fig. \ref{fig:TauM} the cycle period of the cyclic models as a function of $\tau_M$ estimated with Eq. \ref{eq:MaxwellTorqueTime}. We find a positive correlation between the two timescales: if the Maxwell torque timescale is short, the magnetic cycle is short as well. In addition, $\tau_M$ is systematically smaller than $P_{\rm cyc}$ in our simulations, showing that the magnetic feedback is indeed fast enough to act over the magnetic cycle timescale. This further strengthens our interpretation of the non-linear dynamo action occurring in our set of simulations.

Two additional remarkable features can be noted from Fig. \ref{fig:TauM}. First, there seems to be a saturation of the cycle period as $\tau_M$ increases past a threshold. This is not necessarily a surprise: as $\tau_M$ increases, the feedback of the large-scale magnetic field on the differential rotation weakens to ultimately become negligible \modifAS{(or, equivalently, the feedback occurs on a timescale too long compared to the convective turnover timescale)}. As a result, we expect our model to possess a maximal cycle length, leading to such a thresholding effect. Second, one simulation clearly stands out of this $P_{cyc}$--$\tau_M$ relationship (red circle at the bottom of Fig. \ref{fig:TauM}). This particular simulation is model S2a, which is the cyclic solution with the largest Rossby number in our sample ($\mathrm{Ro}_b$ $\sim$ 1, see Table \ref{ta:Params}). This particular model is at the edge of the high-Rossby number transition and starts to exhibit an anti-solar differential rotation profile. \modifAS{It is well known that magnetic torques and stresses can affect the delicate balance establishing the differential rotation profile \citep[\textit{e.g.}][]{Brun:2004ji,Gastine:2014jr,Fan:2014ct,Karak:2015dw}. As a result, models lying very close to the Rossby transition (such as model S2a) are not necessarily expected to fall on the same trend as the other cyclic models in our sample. \modifPC{Nonetheless,}
model S2a still follows the robust Rossby number trend identified in Fig. \ref{fig:RossbyTrend}.}


\modifAS{Several of our models do not produce a deep-seated magnetic cycle (see Fig. \ref{fig:SummaryPlot}), which is found to 
\modifPC{materialize}
only in a relatively narrow Rossby number range. A sharp transition is
observed at $\mathrm{Ro}_b \sim 0.25$ and $\mathrm{Ro}_b \sim 1.0$,
where the Maxwell torque timescale diverges and becomes too long to
\reff{participate in setting the period} of the deep-seated cycle. We now detail what occurs at both of these transitions.} 

\begin{figure*}[htb]
  \centering
  \includegraphics[width=\linewidth]{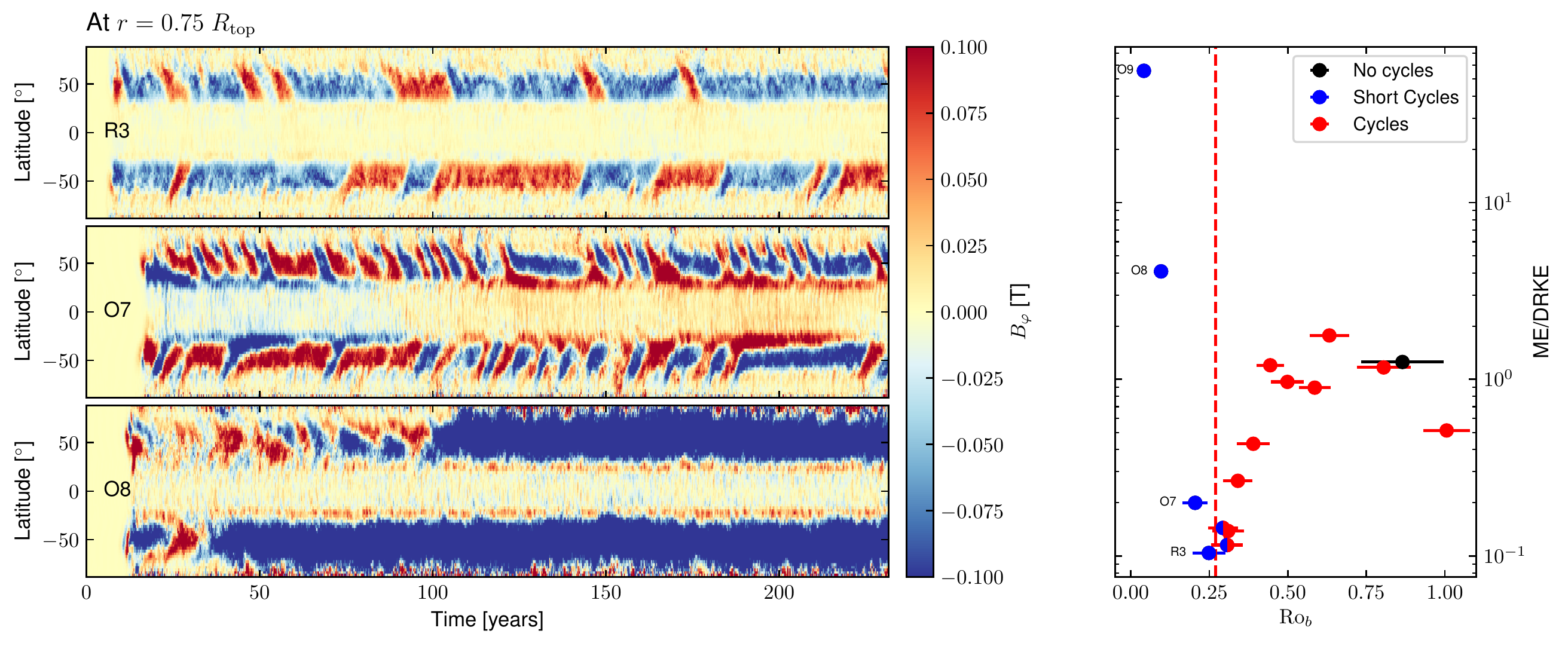}
  \caption{\textit{Left panels.} Time-latitude diagrams of the longitudinally-averaged azimuthal magnetic field at $r=0.75\, R_{\rm top}$, for models R3, O7 and O8 (from top to bottom) spanning the low-Rossby number transition. \textit{Right panel.} Total magnetic energy ME (Eq. \ref{eq:ME} normalized to the convective kinetic energy CKE (Eq. \ref{eq:CKE}) as a function of the Rossby number. As in Fig. \ref{fig:SummaryPlot}, cyclic models are indicated in red and non-cyclic models in black. The vertical red dashed line indicates the Rossby number threshold under which no deep-seated magnetic cycle is found.}
  \label{fig:LowRo}
\end{figure*}

\subsection{Low Rossby number regime}
\label{sec:low-rossby-number}

As we increase the rotation rate of the convective envelope or its density contrast, the Rossby number decreases (Table \ref{ta:Params}). We observe that when $\mathrm{Ro}_b\lesssim 0.25$ the dynamo loses its deep-seated cycle (see Table \ref{ta:cycles} and Fig. \ref{fig:SummaryPlot}). We display in the left panels of Fig. \ref{fig:LowRo} three cases (R3, O7 and O8) tracing the transition of our dynamo solutions from a cyclic solution ($\mathrm{Ro} > 0.25$), to a randomly reversing dynamo (cases R3 and O7), to solutions developing stable wreaths of toroidal field at the base of the convective envelope (cases O8 and O9). The latter cases are morphologically similar to the simulations of \citet{Brown:2010cn} who originally showed that strong azimuthal magnetic structures could be sustained within turbulent convective envelopes. 

As the Rossby number decreases, the total magnetic energy strongly increases in proportion of the differential rotation kinetic energy (see right panel in Fig. \ref{fig:LowRo}). In the meanwhile, the azimuthal magnetic field energy strongly decreases relatively to the total magnetic energy (see Table \ref{ta:energetics}). As a result, the fluctuating (\textit{i.e.} non-axisymmetric) magnetic energy dominates the total magnetic energy for these models. The large-scale feedback from the Maxwell torque is thus much less efficient in these cases, which is why the dynamo does not operate in the same way as for $\mathrm{Ro}_b \gtrsim 0.25$. 

\begin{figure}[htb]
  \centering
  \includegraphics[width=\linewidth]{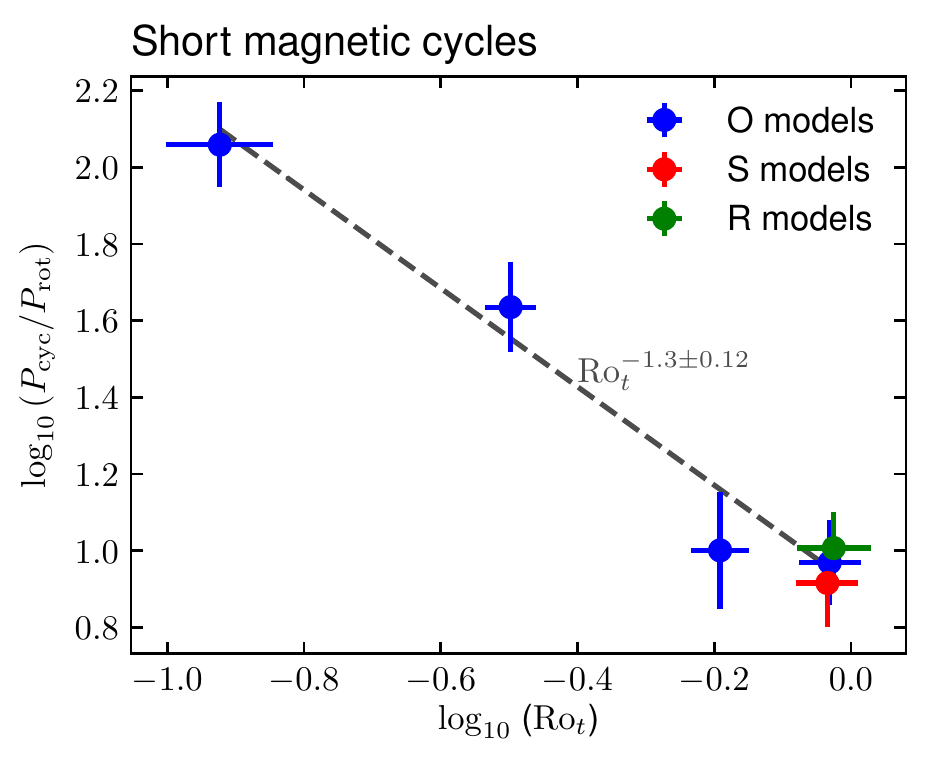}
  \caption{Short cycle period normalized to the stellar rotation period as a function of the Rossby number $\mathrm{Ro}_t$ at the top of the domain, in logarithmic scale. The models are color-coded as in Fig. \ref{fig:RossbyTrend}. Orthogonal distance regression was performed to find the power-law fitted indicated by the gray dashed line, using the error-bars for both the estimated cycle period (Table \ref{ta:cycles}) and Rossby number (Table \ref{ta:Params}).}
  \label{fig:RossbyTrendSec}
\end{figure}


\modifAS{\modifPC{Simulations operating near}
the low-Rossby number transition 
develop a short\modifPC{-period magnetic} cycle (blue circles in the right panel of Fig. \ref{fig:LowRo}) near the top of the domain. We display the short cycle period normalized to the rotation rate of the model as a function of the Rossby number $\mathrm{Ro}_t$ in Fig. \ref{fig:RossbyTrendSec}. Even though we only have 6 models exhibiting such a short cycle, a clear decreasing trend with the local Rossby number is also observed (compare to the deep-seated magnetic cycle in Fig. \ref{fig:RossbyTrend}). The power-law is shallower for the short cycle \modifASt{(keeping in mind that the exact slope value is strongly influenced by a few simulations in the sample)}, but is still characterized by a slope steeper than $-1$, which suggests a trend similar to the deep-seated cycle. Similarly to \citet{Kapyla:2014jr}, we find that a stronger stratification favors such short cycles near the top of the domain (compare, \textit{e.g.}, models R2 and R3), which here simply reflects a decrease of the local Rossby number $\mathrm{Ro}_t$ when the number of density scale heights embedded in the model is increased. Interestingly, the short cycles are observed as soon as $\mathrm{Ro}_t$ is smaller than one, and we see no hint of a short-cycle disappearance at low $\mathrm{Ro}_t$.}

\subsection{Large Rossby number regime}
\label{sec:large-rossby-number}

\begin{figure}[htb]
  \centering
  \includegraphics[width=\linewidth]{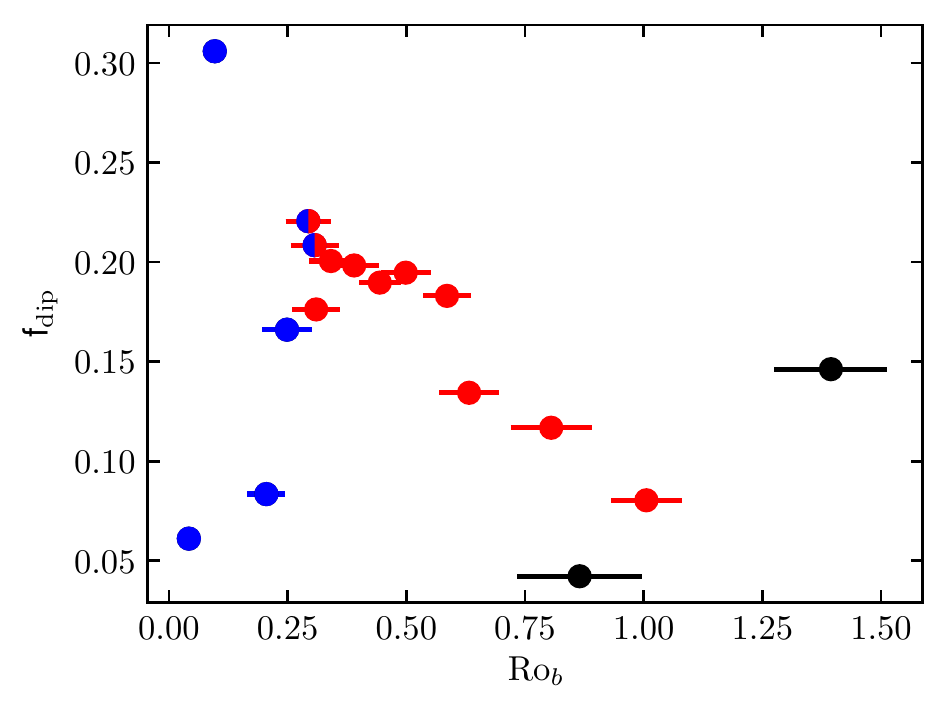}
  \caption{Strength of the dipole ($f_{\rm dip}$) as function of the Rossby number $\Ro_b$ in the $r\in [0.75\,R_{\rm top}, 0.8\,R_{\rm top}]$ interval. The layout is the same as in the right panel of Fig. \ref{fig:LowRo}.}
  \label{fig:fdip}
\end{figure}

\modifAS{In our set of simulations, the deep-seated magnetic cycles are also lost when the Rossby number exceeds unity. A structural change in the differential rotation occurs when $\mathrm{Ro} \sim 1$ \citep[\textit{e.g.}][]{Brun:2017em}, with the differential rotation switching from solar-like to antisolar (fast to slow equator, slow to fast high latitudes). This transition is observed for instance in panel h of Fig. \ref{fig:3repcases} for model sO1 for which a strong antisolar differential rotation starts to appear. Model sO1 interestingly develops a stable azimuthal wreath at the base of the convection zone, akin to the low-Rossby number cases. In this case, though, the differential rotation and convective energies are balanced, and the magnetic energy is less than 10\% of the total kinetic energy (see Table \ref{ta:energetics}). The magnetic energy is dominated by the non-axisymmetric components of the field, which points to a decrease of the large-scale field as the Rossby number of our models is increased. This is further confirmed when computing the dipole strength $f_{\rm dip}$, defined here as the energy of the axisymmetric dipole divided by the total magnetic energy, and shown in Fig. \ref{fig:fdip}. We observe that the dipolar strength decreases as a function of the Rossby number $\mathrm{Ro}_b$, and report a hint of saturation at high Rossby number around $f_{\rm dip}\sim 0.1-0.2$. \modifASt{Additional models at higher Rossby numbers are nevertheless required to fully investigate this trend in the context of the effect of a dynamo transition for old, slowly rotating stars \citep[e.g.][]{Metcalfe:2017kx}, which we leave here for future work.}}

\section{Conclusions}
\label{sec:conclusions}

In this paper, we offered a detailed presentation of an extended set of global MHD simulations of solar convection and dynamo action. Generated using the EULAG code, this simulation set extends the subset originally published in \citet{Strugarek:2017go}. Considering a spherical convective shell with a solar-like aspect ratio, we varied the rotation rate of the shell, its convective luminosity and the number of density scale-heights spanning it. We found that, for adequate parameters well characterized by a range in Rossby number (Eq. \ref{eq:Rossby}), the dynamo sustained in the turbulent spherical shell generates a spatially well-organized large-scale magnetic field exhibiting a fairly regular cyclic behavior. When $\mathrm{Ro}_b \lesssim 0.25$, the dynamo loses its regular cycles and irregularly reverses. Stable large-scale magnetic structures resembling the magnetic wreaths found by \citet{Brown:2010cn} are observed when the Rossby number is even further decreased. When $\mathrm{Ro}_b \gtrsim 1$, the differential rotation switches from solar-like (fast equator, slow poles) to antisolar-like (slow equator, fast high latitudes), as in \citet{Brun:2017em}. The cyclic dynamo also disappears, and the magnetic field becomes more and more non-axisymmetric while retaining a predominantly azimuthal component at the base of the convection zone. We recall here that the EULAG code uses an implicit large-eddy simulation (ILES) approach, where all the dissipation is handled by the numerical scheme itself. This has the important consequence that the large spatial scales in this set of simulations are essentially inviscid, and the dissipation at mid- to small scales is approximated by standard Laplacian-like dissipative operators, with dissipation coefficients reported in Table \ref{ta:dissipation} (see also \citealt{Strugarek:2016bk}).

The cyclic dynamo states found in our 3D, non-linear turbulent simulations are in a fundamentally non-linear regime where the feedback of the Lorentz force on the differential rotation plays an essential role in setting the cycle period. The feedback induces variations of only $\sim 1\%$ of the total differential rotation, but is strong enough to locally affect the latitudinal gradient of the differential rotation and trigger the reversal of the global magnetic field. This interpretation, supported by the detailed analysis of dynamo action in \S\ref{sec:dynamo-action}, is further confirmed by the correlation between the cycle period and the characteristic timescale associated with the Maxwell torques (\S\ref{sec:trends-cycl-solut}).

\modifASt{If the dynamo action observed in this series of simulations does occur in the convective envelope of solar-like stars, several important conclusions for stellar cycles can be drawn from this study. 

First, the cycle period of such stars is set by the non-linear feedback of Maxwell torque deep inside the convection zone. As a result, the surface effects such as the Babcock-Leighton mechanism, while certainly possible and in fact observed at the solar surface, are not essential to the dynamo nor to its cyclic character. 

Second, close to the low-Rossby transition, two types of cycles can easily be mixed when comparing observational results. These cycles, which are non-linearly coupled in turbulent convective envelopes, are not necessarily expected to follow the same trends and dependencies, nor even to originate from the exact same dynamo mechanism. Indeed, in our set of simulations we find two different trends for the magnetic cycle periods, and we also find different propagation properties of the cyclic fields (see panels d1 and d2 in Fig. \ref{fig:3repcases}). Particular care is thus in order when determining trends (or branches) from observed cycles on a large sample of stars. 

Third, at high Rossby numbers, our results suggest a
change in the dynamo state promoting smaller scale, less
axisymmetric fields. This transition at $\Ro\sim 1$ is interesting in several aspects.
First it has been shown to likely be the transition of the state of differential rotation from
solar to anti-solar (\citealt{Brun:2017em} and references therein). Second, it has been argued by \citep{vanSaders:2016cr,Metcalfe:2017kx} 
that such a change could impact the type of dynamo occurring (by modifying, \textit{e.g.} the omega-effect) and hence explain
a possible break of stellar rotation-age relationship (so-called gyrochronology, see \citealt{Barnes:2003ga}).
In our set of simulations, at the transition $\Ro\sim 1$ the large-scale field does not disappear and the dipolar fraction
remains within 10$\%$ of the total magnetic energy. Hence the change of the properties of the dynamo field we observe at $\Ro\sim 1$ 
does not seem to be large enough to explain by itself a strong decrease of the torque exerted by the magnetized wind
of the star. We intend to explore further dynamo action near the $\Ro\sim 1$ transition to
quantitatively assess how the dynamo and magnetic activity is modified and how this affect the stellar wind properties.}

In the regime where a cyclic dynamo is found, \citet{Strugarek:2017go} showed that the solar cycle period fits very well the Rossby trend we found. However, some solar-type stars of the analyzed sample diverged significantly from our single-branch theoretical trend. These differences could be ascribed to varying aspect ratios and absolute thickness of the convection zone, thus spanning a different range of density scale heights. The differential rotation and Rossby number achieved at the base of such convection zones are thus expected to have different dependency to the stellar parameters (rotation, luminosity) than in our study where we kept the aspect ratio constant. Other stellar internal structures (as in \citealt{Brun:2017em}), eventually including the coupling with an underlying stable layer (see \citealt{Beaudoin:2018tl} for a first step on this aspect), are now needed to assess the robustness of our scaling-law for solar-like stars of different spectral types and metallicities.

Finally, we want to stress again that the simulations presented in this study operate in an overall parameter regime very far from that expected to
characterize physical conditions in solar and stellar interiors. \refff{Furthermore, a systematic comparison of the different ways of forcing convection in stellar envelopes and
how surface layers and stratification may influence the description of stellar turbulent transport of
heat, angular momentum and magnetic field is in order in the
community, and some attempts in these directions have been initiated \citep[see \textit{e.g.}][]{Lord:2014hu,Strugarek:2016bk,2016ApJ...829L..17C,Cossette:2017kf,2018ApJ...859..117N}.}
 The dynamo mechanism identified in this series of simulations nevertheless gives a novel robust mechanism explaining the origin of magnetic cycles in turbulent convective envelopes. These simulations are able to reproduce the solar cycle period, in spite of our relatively coarse spatial discretization grid. We hope in a near future that better resolved simulations, and ultimately comparison with results obtained with other numerical techniques, will help enhance the realism of the dynamo mechanism unveiled in this study, and further validate it applicability to stellar dynamo and cycles.

\acknowledgments

We thank S. Tobias and J.-D. do Nascimento Jr. for useful discussions
on magnetic cycles and dynamos, \reff{and an anonymous referee for
  suggestions that improved the presentation of our results}.
The authors acknowledge support from the Natural Sciences and Engineering Research Council of Canada. A. Strugarek acknowledges support from CNES in preparation of the Solar Orbiter mission. This work was also supported by the INSU/PNST. We acknowledge access to super-computers through GENCI (project 1623) and Calcul Qu\'ebec, a member of the Compute Canada consortium.

\bibliographystyle{yahapj}

\appendix

\section{Definitions of the global energies}
\label{sec:DefEnergies}

\modifAS{We quantify the energies of each simulation by defining the total kinetic energy (KE), along with its differential rotation (DRKE) and convective (CKE) components, as follows
\begin{eqnarray}
	\label{eq:KE}
\mbox{KE} &=& \left\langle \iiint \frac{1}{2}\bar{\rho} \left(u_r^2 + u_\theta^2 + u_\varphi^2\right) {\rm d}V \right\rangle_t \, , \\
	\label{eq:DRKE}
\mbox{DRKE} &=& \left\langle \iiint \frac{1}{2}\bar{\rho} \left\langle u_\varphi\right\rangle_\varphi ^2 {\rm d}V \right\rangle_t \, , \\
	\label{eq:CKE}
\mbox{CKE} &=& \left\langle \iiint \frac{1}{2}\bar{\rho} \left( \tilde{u}_r^2 + \tilde{u}_\theta^2 + \tilde{u}_\varphi^2\right)  {\rm d}V \right\rangle_t \, , 
\end{eqnarray}
where $\left\langle\right\rangle_t$ stands for the temporal average, $\left\langle\right\rangle_\varphi$ for the azimuthal average, $\tilde{X} = X - \left\langle X \right\rangle_\varphi$ and ${\rm d}V = r^2\sin\theta {\rm d}r {\rm d}\theta {\rm d}\varphi $. We furthermore define the total magnetic energy (ME) and its mean toroidal (TME), mean poloidal (PME) and fluctuating (FME) components as 
\begin{eqnarray}
	\label{eq:ME}
\mbox{ME} &=& \left\langle \iiint \frac{1}{2\mu_0} \left(B_r^2 + B_\theta^2 + B_\varphi^2\right) {\rm d}V \right\rangle_t \, , \\
	\label{eq:TME}
\mbox{TME} &=& \left\langle \iiint \frac{1}{2\mu_0} \left\langle B_\varphi\right\rangle_\varphi ^2 {\rm d}V \right\rangle_t \, , \\
	\label{eq:PME}
\mbox{PME} &=& \left\langle \iiint \frac{1}{2\mu_0} \left( \left\langle B_r\right\rangle_\varphi^2 + \left\langle B_\theta\right\rangle_\varphi^2 \right) {\rm d}V \right\rangle_t \, , \\
	\label{eq:FME}
\mbox{FME} &=& \left\langle \iiint \frac{1}{2\mu_0} \left( \tilde{B}_r^2 + \tilde{B}_\theta^2 + \tilde{B}_\varphi^2\right)  {\rm d}V \right\rangle_t \, . 
\end{eqnarray}
All these energies are listed in Table \ref{ta:energetics} for the whole simulation set.}

\section{Dissipative properties of the simulations}
\label{sec:DissipativeProp}

\begin{figure}[htb]
  \centering
\includegraphics[width=0.8\linewidth]{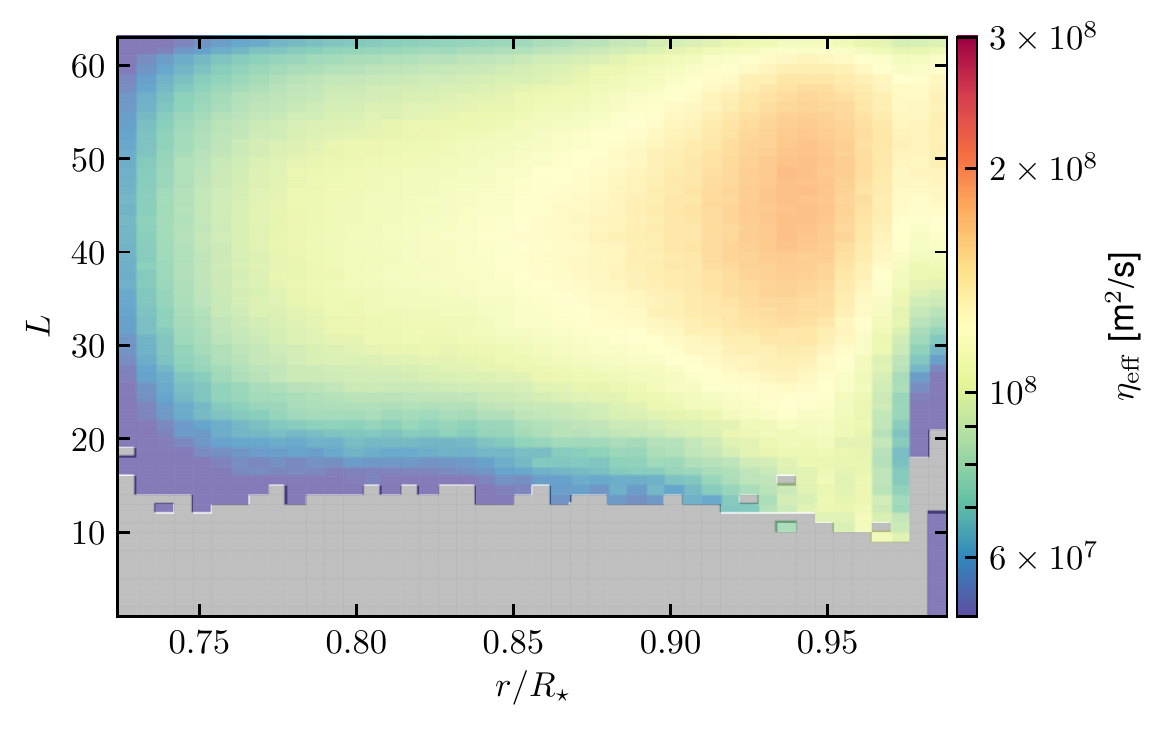}
  \caption{Effective ohmic dissipation coefficient $\eta_{\rm eff}$ as a function of the spherical harmonic degree $L$ and depth of the convection zone. The grey area labels scales and depths where the numerical dissipation is weak and cannot be approximated by a classical laplacian operator.}
  \label{fig:etaeff}
\end{figure}

\modifAS{
We characterize the dissipative properties of the models following the methodology detailed in \citet{Strugarek:2016bk}. The basic principle is as follows: the set of anelastic equations (\ref{eq:LBR}-\ref{eq:LBR_B}) is projected on the spherical harmonics basis. By projecting each spectral equation (for example, we project Equation \ref{eq:LBRET_u} on the vectorial spherical harmonics and take the dot-product with $\mathbf{u}$) and integrating them over concentric spheres, we obtain evolution equations for the kinetic energy, potential temperature, and magnetic energy spectra (see \citealt{Strugarek:2016bk}). The evolution equation for the magnetic energy spectrum
\begin{equation}
	\mathcal{E}_L^M(r) = \frac{1}{2\mu_0} \iint {\bf B}_L \cdot {\bf B}_L^{cc} {\rm d\Omega}
\end{equation}
can be written, using the same notation as in \citealt{Strugarek:2016bk},
\begin{equation}
	\dot{\mathcal{E}}^M_L(r) = \mathcal{T}_L(r) \, ,
\end{equation}
where $\mathcal{T}_L$ is the energy transfer from the kinetic energy reservoir (see \citealt{Strugarek:2013kt}). Formally, EULAG-MHD solves the induction equation with no ohmic dissipation. As a result, the residual $\dot{\mathcal{E}}^M_L - \mathcal{T}_L $ is an estimate of the level of numerical dissipation added by the MPDATA algorithm, which is generically scale and time-dependent. We match here this residual (averaged over time) to an explicit ohmic dissipation with an effective coefficient $\eta_{\rm eff}$, namely 
\begin{equation}
\mathcal{O}_L = -\iint \left.\boldsymbol{\nabla}\times\left( \eta_{\rm eff} \boldsymbol{\nabla}\times{\bf B}\right)\right|_L \cdot {\bf B}_L^{cc} {\rm d\Omega}\, .
\end{equation}
The procedure was illustrated at length in \citet{Strugarek:2016bk}, we do not repeat it here and simply present in Fig. \ref{fig:etaeff} the resulting $\eta_{\rm eff}$ as a function of depth and spherical harmonic degree $L$ for model O5. We see that the effective ohmic dissipation is of the order of 1-2$\times$10$^{8}$ m$^2/s$ in the bulk of the convective envelope for modes $L > 25$. The grey area corresponds to a region where the effective dissipation cannot be matched to a classical laplacian operator and is smaller than the dissipation at larger scales. The spatial distribution with slightly higher $\eta_{\rm eff}$ in the upper part of the convection zone is a direct consequence of the scale distribution of the magnetic structures as a function of depth. We note one more time that the largest scales in the domain (typically $L<10$) are almost not dissipated compared to the smallest scales, which are in turn the scales well characterized by the dissipation coefficients derived through this methodology.

Finally, the overall procedure is repeated for all models and also for the effective viscosity $\nu_{\rm eff}$ and heat dissipation coefficient $\kappa_{\rm eff}$, which are given in Table \ref{ta:dissipation}. 
}

\begin{deluxetable}{lccccc}
  \tablecaption{Effective dissipation at small scales \label{ta:dissipation}}
  \tablecolumns{6}
  \tabletypesize{\scriptsize}
  \tablehead{
    \colhead{Case}  &
    \colhead{$\nu_{\rm eff}$} &
    \colhead{$\kappa_{\rm eff}$} &
    \colhead{$\eta_{\rm eff}$} &
    \colhead{$P_r$} &
    \colhead{$P_m$} 
\\
   &
   \colhead{[$10^{8}$ m$^2$/s]} &
   \colhead{[$10^{8}$ m$^2$/s]} &
   \colhead{[$10^{8}$ m$^2$/s]} &
   \colhead{} &
   \colhead{} 
  }
  \startdata 
\input{DissipationTable.tex}
  \enddata
\end{deluxetable}

\section{Empirical Mode Decomposition algorithm}
\label{sec:emd}

\begin{figure}[htb]
  \centering
\includegraphics[width=\linewidth]{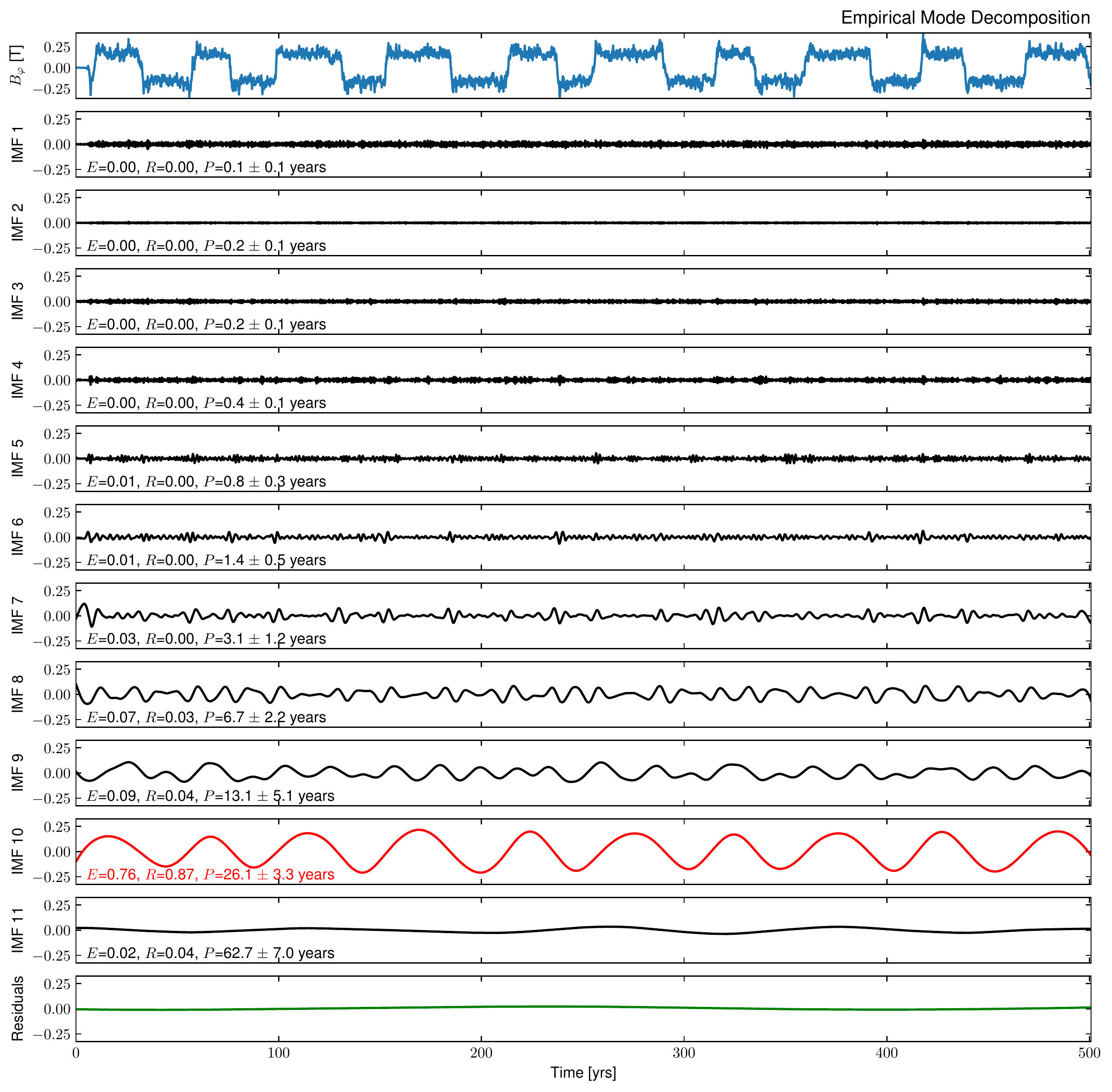}
  \caption{Empirical mode decomposition of $\left\langle B_\varphi \right\rangle_\varphi$ at $r=0.72\, R_\odot$ and latitude $57^\circ$ for model O5. The top panel (blue) shows the original signal as a function of time, panels labeled IMF 1 to IMF 11 show the intrinsic mode functions (IMFs) of the signal decomposed with the CEEMDAN method, and the green line at the bottom shows the residual of the decomposition. In each panel, the energy ($E$), robustness ($R$) and period ($P$) of the displayed IMF is indicated in the bottom left.}
  \label{fig:ceeemdan}
\end{figure}

\modifAS{In order to robustly extract the periodicity of the dynamo cycle, we make use in this work the Complete variant of the Ensemble Empirical Mode Decomposition with Adaptative Noise (CEEMDAN algorithm). Empirical Mode Decomposition (EMD) is a method to decompose an input signal into a series of Intrinsic Mode Functions (IMFs), which are oscillatory modes that are not required to be simple sinusoidal functions but still retain meaningful local frequencies (for an in-depth discussion of the EMD method, see \citealt{Luukko:2015jy,Kapyla:2016hg} and references therein). The CEEMDAN variant was further developed by \citet{Torres:2011bo}, which achieves completeness of the decomposition and improves the robustness of the method for noisy signals. The CEEMDAN method has been made available in the open-source libeemd library \citep{Luukko:2015jy}, which was used in this work. We show in Fig. \ref{fig:ceeemdan} one example of a CEEMDAN run applied to $\left\langle B_\varphi \right\rangle_\varphi$ at $r=0.72\, R_\odot$ and latitude $57^\circ$ for model O5. The original signal is shown at the top panel in blue, in physical units. Given the length, mean and standard deviation of the signal, the CEEMDAN method decomposes it automatically into 11 IMFs that are shown in the other panels. The residual of the decomposition is shown in green in the bottom panel. For each IMF, one can compute its relative energy $E$ to assess how much its contributes to the total signal. In this particular example, IMF 10 (in red) completely dominates the decomposition with a relative energy of 76\%. It is also important to assess the significance of the decomposition, characterized here by the robustness parameter $R$ indicated in each panel. To have an estimate of the robustness of each IMF, we build a white noise signal of the same length as the original signal with the same mean and standard deviation. We perform the CEEMDAN on this white noise signal, and calculate the robustness of each IMF of the original signal by its relative energy divided by the relative energy of its white noise IMF counterpart. The parameter $R$ is then the normalized so that the sum of the robustnesses is 1, and we find that IMF 10 is also very robust (87\%) giving us confidence that it captures accurately the main periodicity of the original signal. Finally, a mean period (P) with its error-bars can be easily computed for each IMF, giving us a robust estimate of the main periodicity of the dynamo at that depth and latitude for model O5.}

\end{document}

%% file: EnergyTable.tex
sO1 &  $20.8 \pm $4.0 &  $42.0 \pm $18.4 &  $56.2 \pm $4.4 &     $7.1 \pm $4.1 &  $21.6 \pm $14.3 &   $4.9 \pm $3.1 &  $73.5 \pm $41.2 \\
O1  &  $12.1 \pm $0.4 &   $12.3 \pm $1.9 &  $86.9 \pm $2.9 &    $13.9 \pm $4.0 &   $24.9 \pm $8.2 &   $8.7 \pm $2.5 &  $66.4 \pm $18.9 \\
O2  &  $10.7 \pm $0.2 &    $9.2 \pm $1.2 &  $90.1 \pm $2.2 &    $16.0 \pm $4.1 &   $24.1 \pm $7.1 &  $12.6 \pm $2.9 &  $63.2 \pm $16.2 \\
O3  &  $11.0 \pm $0.7 &   $18.9 \pm $5.8 &  $80.7 \pm $2.3 &    $15.2 \pm $6.0 &  $22.2 \pm $10.5 &  $17.0 \pm $5.8 &  $60.8 \pm $23.7 \\
O4  &  $13.1 \pm $1.2 &   $38.5 \pm $8.9 &  $61.2 \pm $1.7 &    $14.2 \pm $7.4 &  $22.6 \pm $15.8 &  $19.4 \pm $8.4 &  $58.0 \pm $28.3 \\
O5  &  $14.7 \pm $1.4 &   $48.3 \pm $9.0 &  $51.5 \pm $1.5 &    $11.4 \pm $5.9 &  $20.7 \pm $15.5 &  $21.3 \pm $9.3 &  $57.9 \pm $27.9 \\
O6  &  $17.0 \pm $1.2 &   $58.1 \pm $7.0 &  $41.7 \pm $1.2 &     $7.9 \pm $3.3 &  $16.8 \pm $11.4 &  $23.4 \pm $8.6 &  $59.8 \pm $22.8 \\
O7  &  $15.4 \pm $1.4 &   $57.1 \pm $9.1 &  $42.7 \pm $1.5 &    $11.1 \pm $1.7 &   $12.8 \pm $4.9 &  $14.6 \pm $4.6 &  $72.5 \pm $12.1 \\
O8  &   $6.7 \pm $0.3 &   $31.1 \pm $5.6 &  $68.8 \pm $1.8 &  $121.0 \pm $20.6 &   $17.8 \pm $5.8 &  $14.8 \pm $4.3 &   $67.4 \pm $8.4 \\
O9  &   $2.7 \pm $0.1 &    $9.3 \pm $2.1 &  $90.6 \pm $2.5 &  $497.2 \pm $58.4 &    $5.8 \pm $2.9 &   $5.0 \pm $2.5 &   $89.3 \pm $9.1\vspace*{0.1cm}\\
S1  &  $14.3 \pm $0.8 &   $58.9 \pm $5.5 &  $41.0 \pm $1.3 &     $6.6 \pm $2.1 &   $14.8 \pm $5.9 &  $24.2 \pm $7.4 &  $61.1 \pm $19.0 \\
S2a &  $20.2 \pm $0.8 &   $16.5 \pm $3.0 &  $82.5 \pm $2.9 &     $8.4 \pm $3.4 &  $24.5 \pm $11.8 &   $5.7 \pm $2.1 &  $69.8 \pm $27.0 \\
S2b &  $15.8 \pm $1.1 &   $20.7 \pm $7.3 &  $79.0 \pm $2.0 &    $19.4 \pm $8.3 &  $26.7 \pm $14.8 &  $16.6 \pm $5.9 &  $56.7 \pm $22.6\vspace*{0.1cm}\\
R0  &  $41.8 \pm $0.7 &    $8.7 \pm $1.0 &  $91.0 \pm $1.8 &    $10.6 \pm $2.0 &   $27.1 \pm $3.5 &  $15.7 \pm $3.6 &  $57.2 \pm $12.6 \\
R1  &  $24.3 \pm $1.6 &   $20.4 \pm $6.1 &  $79.2 \pm $1.8 &    $15.3 \pm $6.0 &  $26.5 \pm $11.3 &  $17.1 \pm $6.2 &  $56.3 \pm $22.2 \\
R2  &  $13.2 \pm $0.6 &   $54.9 \pm $4.5 &  $44.9 \pm $1.3 &     $7.4 \pm $2.2 &   $15.1 \pm $5.5 &  $24.7 \pm $6.6 &  $60.2 \pm $18.5 \\
R3  &   $9.8 \pm $0.4 &   $59.5 \pm $3.9 &  $40.3 \pm $1.5 &     $6.1 \pm $0.7 &   $11.7 \pm $1.9 &  $21.8 \pm $4.5 &   $66.6 \pm $9.1

%% file: Cycles.tex
sO1 &                 - &      - &      - &                - &      - &      - \\
O1  &  $13.10 \pm 2.50$ & $0.63$ & $0.72$ &                - &      - &      - \\
O2  &  $16.47 \pm 2.84$ & $0.72$ & $0.89$ &                - &      - &      - \\
O3  &  $15.97 \pm 2.69$ & $0.82$ & $0.92$ &                - &      - &      - \\
O4  &  $21.89 \pm 1.91$ & $0.79$ & $0.97$ &                - &      - &      - \\
O5  &  $26.01 \pm 2.79$ & $0.65$ & $0.89$ &                - &      - &      - \\
O6  &  $31.40 \pm 6.21$ & $0.74$ & $0.87$ &  $0.54 \pm 0.16$ & $0.38$ & $0.26$ \\
O7  &                 - &      - &      - &  $0.43 \pm 0.18$ & $0.61$ & $0.58$ \\
O8  &                 - &      - &      - &  $1.03 \pm 0.32$ & $0.51$ & $0.48$ \\
O9  &                 - &      - &      - &  $1.49 \pm 0.43$ & $0.41$ & $0.60$\vspace*{0.1cm} \\
S1  &  $32.27 \pm 8.51$ & $0.58$ & $0.83$ &  $0.53 \pm 0.16$ & $0.39$ & $0.29$ \\
S2a &   $8.88 \pm 1.74$ & $0.57$ & $0.85$ &                - &      - &      - \\
S2b &  $17.52 \pm 4.18$ & $0.64$ & $0.76$ &                - &      - &      -\vspace*{0.1cm} \\
R0  &                 - &      - &      - &                - &      - &      - \\
R1  &  $16.80 \pm 2.84$ & $0.84$ & $0.97$ &                - &      - &      - \\
R2  &  $29.88 \pm 4.13$ & $0.61$ & $0.88$ &                - &      - &      - \\
R3  &                 - &      - &      - &  $0.66 \pm 0.16$ & $0.70$ & $0.64$ 

%% file: DissipationTable.tex
sO1 & $1.12$ & $0.78$ & $1.16$ & $1.44$ & $0.97$ \\
O1  & $1.30$ & $0.84$ & $1.11$ & $1.55$ & $1.17$ \\
O2  & $1.37$ & $0.92$ & $1.11$ & $1.49$ & $1.24$ \\
O3  & $1.40$ & $1.07$ & $1.13$ & $1.31$ & $1.24$ \\
O4  & $1.37$ & $1.18$ & $1.10$ & $1.16$ & $1.25$ \\
O5  & $1.41$ & $1.21$ & $1.04$ & $1.17$ & $1.35$ \\
O6  & $1.39$ & $1.21$ & $1.03$ & $1.15$ & $1.35$ \\
O7  & $1.59$ & $1.19$ & $0.85$ & $1.33$ & $1.87$ \\
O8  & $2.06$ & $0.74$ & $0.95$ & $2.79$ & $2.18$ \\
O9  & $2.22$ & $0.45$ & $0.50$ & $4.95$ & $4.46$\vspace*{0.1cm} \\
S1  & $1.20$ & $1.12$ & $0.91$ & $1.07$ & $1.32$ \\
S2a & $1.55$ & $0.49$ & $1.30$ & $3.19$ & $1.19$ \\
S2b & $1.81$ & $0.94$ & $1.39$ & $1.92$ & $1.30$\vspace*{0.1cm} \\
R0  & $2.36$ & $1.85$ & $1.60$ & $1.27$ & $1.48$ \\
R1  & $2.09$ & $1.56$ & $1.58$ & $1.34$ & $1.33$ \\
R2  & $1.22$ & $1.09$ & $0.92$ & $1.12$ & $1.33$ \\
R3  & $1.02$ & $0.88$ & $0.71$ & $1.16$ & $1.44$ \\